\documentclass[%
 amsmath,amssymb, 
 aps,prx,twocolumn,superscriptaddress
]{revtex4-1}
\usepackage[utf8]{inputenc}
\usepackage[T1]{fontenc}
\usepackage{physics}
\usepackage{amsmath}
\usepackage{bbm}
\usepackage{tikz}
\usepackage{subcaption}
\usepackage{float}

\usepackage{color}

\newcommand{\godd}[1]{\gamma_{2{#1}-1}}
\newcommand{\gev}[1]{\gamma_{2{#1}}}

\begin{document}
\title{Custom fermionic codes for quantum simulation}
\author{Riley W. Chien}
\email{Riley.W.Chien.gr@dartmouth.edu}
\affiliation{Department of Physics and Astronomy, Dartmouth College, Hanover, New Hampshire 03755}

\author{James D. Whitfield}
\email{James.D.Whitfield@dartmouth.edu}
\affiliation{Department of Physics and Astronomy, Dartmouth College, Hanover, New Hampshire 03755}

\begin{abstract}
    Simulating a fermionic system on a quantum computer requires encoding the anti-commuting fermionic variables into the operators acting on the qubit Hilbert space. The most familiar of which, the Jordan-Wigner transformation, encodes fermionic operators into non-local qubit operators. As non-local operators lead to a slower quantum simulation, recent works have proposed ways of encoding fermionic systems locally. In this work, we show that locality may in fact be too strict of a condition and the size of operators can be reduced by encoding the system quasi-locally. We give examples relevant to lattice models of condensed matter and systems relevant to quantum gravity such as SYK models. Further, we provide a general construction for designing codes to suit the problem and resources at hand and show how one particular class of quasi-local encodings can be thought of as arising from truncating the state preparation circuit of a local encoding. We end with a discussion of designing codes in the presence of device connectivity constraints.
\end{abstract}

\maketitle

\section{Introduction}

A well-known duality between spins and fermions due to Jordan and Wigner \cite{jordanwigner} has been famously employed in the solutions of spin chains \cite{lieb1961two}. Recent years have seen a new applications of spin-fermion dualities as a way of encoding systems of indistinguishable fermions into a system of distinguishable qubits. Such transformations are employed in the simulation of fermionic systems on quantum computers. The idea of a quantum simulator was conceived in \cite{feynman1999simulating} and further expanded upon in \cite{lloyd1996universal}. It is now expected that quantum computers will become an invaluable tool in the study of physical properties of strongly correlated systems which are out of reach of classical computers such as for example quantum chemistry \cite{mcardle2020quantum}. Target problems include chemical reaction mechanisms \cite{reiher2017elucidating} and the Hubbard model \cite{abrams1997simulation}.

Working in second quantization, it is necessary to encode the anti-commuting nature of the fermionic operators into the local qubit degrees of freedom. The solution used in the Jordan-Wigner transformation (JW), is to map local operators on one side of the duality to non-local operators on the other side. When mapping from fermions to spins, the Pauli Z strings are non-local
\begin{align}
    a_j &\to \prod_{k<j} Z_k (X_j + i Y_j)\\
    a_j^{\dag} &\to \prod_{k<j} Z_k (X_j - i Y_j).
    \label{eq: jw}
\end{align}
In particular, the JW strings can lead to operators as large as the system size.

Such non-local operators are known to lead to larger gate counts in the quantum simulation experiment. In particular, the number of qubits an operator acts on nontrivially, a number we refer to as the Pauli weight, determines the required number of CNOT gates to perform a rotation generated by that operator. Thus, lowering the Pauli weight is an important consideration in particular for the Trotterization paradigm of time evolution which may be used phase estimation,as a component of a variational state preparation step, or in quantum imaginary time evolution \cite{motta2020imaginary}. It is important to mention that in regard to phase estimation, recent work has sought to reduce the T-gate complexity as this is likely to determine the dominant time cost of a simulation given the magic state paradigm of universal fault tolerance \cite{bravyi2005universal}. This work does not seek to address this important issue. Nevertheless, we expect this work to be of broad interest.

In addition to Jordan-Wigner, a number of other mappings that map $N$ fermionic modes to $N$ qubits are known. These include the parity mapping and the Bravyi-Kitaev mapping \cite{bravyi2002fermionic}. Recently, Jiang et al. put forward an encoding based on ternary trees which they show to have an optimal average Pauli weight across all Majorana operators defined on the system \cite{jiang2020optimal}.

Bravyi and Kitaev also proposed in \cite{bravyi2002fermionic} the superfast encoding, an explicit method of encoding a system of fermions into qubit degrees of freedom in such a way as to keep all operators local. This work was closely followed by \cite{ball2005fermions,verstraete2005mapping}. The construction by Bravyi and Kitaev work was explored in \cite{whitfield2016local,setia2018,chien2019analysis} and generalized into the construction we will employ here by Setia et al. in \cite{setia2019superfast}. A number of other encodings which have sought to reduce the necessary qubit resources and Pauli weights of transformed operators have also been proposed \cite{jiang2019majorana,steudtnerAQM}. We mention in particular the low weight encoding of Derby and Klassen \cite{derby2020low} as their construction bears the lowest qubit requirement and operator Pauli weight of known encodings on the square lattice. An investigation of the resulting gauge theory was given by Chen et al. along with a construction that gave more a more careful consideration of spin structures in \cite{chen2018exact,chen2019exact3d}. Local bosonization has also been implemented as a tensor network operator in \cite{shukla2020tensor}, mapping fermionic tensor networks to bosonic tensor networks. Also, recent progress has been made in constructing spin duals of fermionic models in translationally invariant settings using an algebraic formalism \cite{tantivasadakarn2020jordan} with focus towards topological and fracton models.

Understanding these local encodings is best done through the toric code model. This famous model, proposed by Kitaev in \cite{kitaev2003fault}, is a model for a topological quantum memory. For our purposes here, it will be most important to note that there are four basic anyon excitations, the vacuum \textbf{1}, a bosonic charge $e$, a bosonic flux $m$, and a fermionic composite of the charge and flux $\varepsilon = e\times m$. These particles are all created, destroyed, and moved at the endpoints of local string operators, made possible by the fact that the system is topologically ordered \cite{levin2003fermions,levin2005string}. The presence or absence of fermionic anyons at a site on the lattice corresponds to the occupancy of the fermionic mode associated with that site. Finally, all known local fermionic encodings are equivalent to the toric code defined on some lattice up to a constant-depth local Clifford circuit. For example, the Derby-Klassen low weight encoding \cite{derby2020low} uses as its underlying topological state the Wen plaquette model \cite{wen2003quantum} which is equivalent to the toric code up to a local basis change on half of the qubits. Thus, the initialization step of a quantum simulation utilizing a local encoding is equivalent to preparing a toric code state. We will return to this point at times throughout the paper when it becomes important.

In this paper, we first review a general construction for local fermionic encodings before presenting our main result, a generalization of fermionic encodings that is fully customizable to suit the available resources as well as the problem geometry. The encoding presented here also encompasses a number of existing methods as special cases which we mention along the way. We give a number of examples including lattice models relevant to condensed matter physics and highly connected systems relevant to the quantum gravity community which frustrate existing encoding methods. Finally, given that a number of quantum computing architectures are subject to limited connectivity, we discuss designing codes under device connectivity constraints.

\section{Review of local fermionic encoding}

In this section we review a certain construction introduced by Setia et al. in \cite{setia2019superfast} for representing fermionic systems in terms of qubits in a local fashion. This method is reminiscent of the construction of \cite{majorana_dualities} for generating highly entangled states using Majorana modes at ends of nanowires.

Typically, our problem setting will be that we are interested in some property of a system of $N$ fermionic modes with dynamics governed by a Hamiltonian 
\begin{equation}
    H = \sum_{jk} h_{jk} a^{\dag}_{j}a_{k} + \sum_{jklm} h_{jklm} a^{\dag}_{j}a^{\dag}_{k}a_{l}a_{m} + \dots
    \label{eq: fermionic_ham}
\end{equation}
where the fermionic operators satisfy the usual anti-commutation relations $\{a_j, a^{\dag}_k \} = a_j a^{\dag}_k + a^{\dag}_k a_j= \delta_{jk}$ and $\{a_j, a_k \} = \{a^{\dag}_j, a^{\dag}_k \} = 0$. As explored in \cite{zohar2018eliminating}, the Hamiltonian could also couple the fermions to a gauge field, for example in the context of a
high energy physics simulation.

It will be useful to work in the Majorana basis of fermionic operators
\begin{align}
    a_{j} &= \frac{1}{2}(\godd{j} + i\gev{j})\\
    a_{j}^{\dag} &= \frac{1}{2}(\godd{j} - i\gev{j})\\
    \{\gamma_{j},\gamma_{k}\} &= 2\delta_{jk}.
\end{align}
From these, we can form two types of operators quadratic in the Majoranas which we will from here on refer to as edge and vertex operators,
\begin{align}
    A_{jk} &= -i \godd{j}\godd{k} \quad \text{ (edge)}\label{eq:edge_op} \\
    \label{eq:vertex_op}B_j &= -i \godd{j}\gev{j} \quad \text{ (vertex)}. 
\end{align}
These operators suffice to generate the full algebra of parity preserving fermionic operators. The vertex operator $B_j$ is the parity operator for the mode $j$ and the edge operator $A_{jk}$ is involved in all hopping, pairing, and scattering processes. Note that edge operators involve only odd-indexed Majoranas. Appropriately multiplying an edge operator by vertex operators allows for coupling two fermionic modes by their even indexed Majoranas.  For example, a hopping term can be expressed as
\begin{equation}
    a_j^{\dag}a_k + a_k^{\dag}a_j = -i (A_{jk}B_k + B_j A_{jk})/2.
\end{equation}

\begin{figure}
    \begin{subfigure}{.48\textwidth}
    \centering
    \includegraphics[width = 0.6\linewidth]{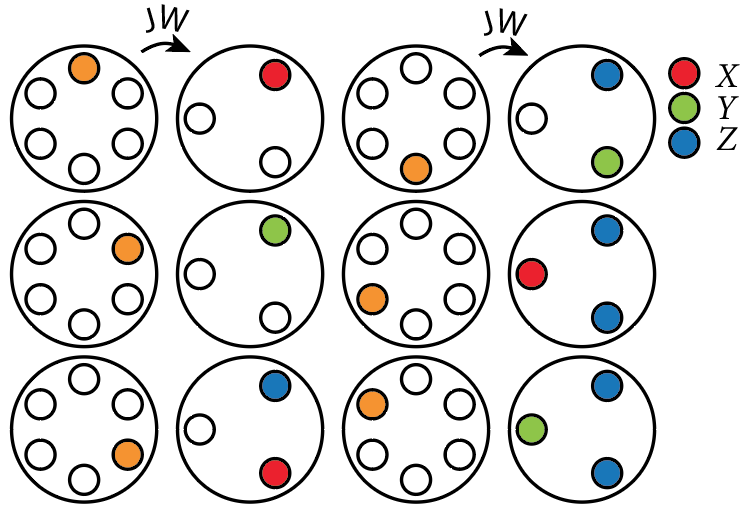}
    \caption{}
    \end{subfigure}
    \begin{subfigure}{0.48\textwidth}
    \centering
    \includegraphics[width = 0.9\linewidth]{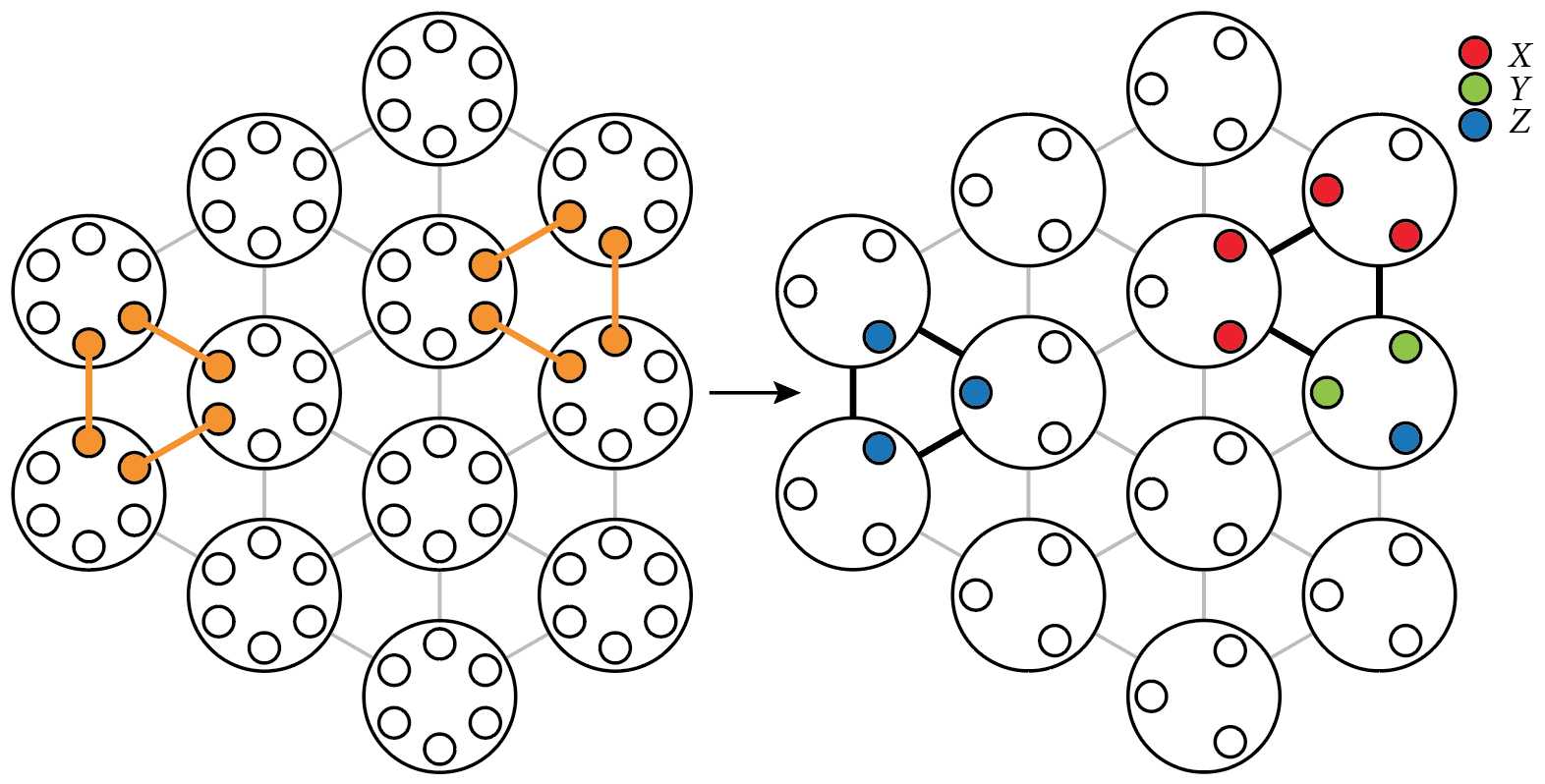}
    \caption{}
    \end{subfigure}
    \caption{(a) Local Majorana modes for a degree $d=6$ vertex. (b) Two types of plaquette stabilizers for the triangular lattice with the above encoded Majoranas. (Left) Orange lines connecting Majoranas indicate the corresponding edge operators. (Right) The corresponding qubit operators.}
    \label{fig:triangular_GSE}
\end{figure}

We here explicitly place the fermionic system on a graph which we refer to as the interaction graph with fermionic modes associated to vertices $j\in V$. When one writes the fermionic Hamiltonian in terms of edge and vertex operators, if an edge operator, $A_{jk}$, coupling modes $j,k$ is required to write the Hamiltonian in this form, then an edge is placed on the graph connecting vertices $j$ and $k$, $(j,k)\in E$. The graph is the given as the pair $\Gamma = (V,E)$. Further, we can identify the set of plaquettes on the graph $P$ and the boundary of a given plaquette $p$ is the set of edges which form it, $\partial p$.

For each vertex $j$ on the graph, we place a number of qubits equal to half the degree of the vertex, $d(j)/2$ (actually $\lceil d(j)/2 \rceil$ but we will assume graphs of even degree for simplicity). We next define $d(j)$ Pauli operators acting on the $d(j)/2$ qubits on site $j$ which we will refer to as local Majoranas
\begin{equation}
    c_{j}^{1},\dots,c_{j}^{d(j)}\in \mathcal{P}_{d(j)/2}
    \label{eq:local_majoranas}
\end{equation}
We use the letter $c$ for the local Majoranas to distinguish these Pauli operators from the physical Majorana fermion operators $\gamma$. These local Majoranas may be anything except that they must satisfy the following properties: (1) they must satisfy Majorana anti-commutation relations with the other operators defined on that vertex (they commute with operators defined on other vertices) and (2) they must generate the full Pauli group on $d(j)/2$ qubits. Any choice of definition for the local Majoranas is related to any other choice by a Clifford circuit acting only on qubits at that vertex. In the figures throughout this paper, we will always use Jordan-Wigner to encode the local Majoranas on a given vertex starting from the top clockwise, $\{c_j^1,c_j^2,c_j^3,c_j^4,\dots\} \to \{X_{j1},Y_{j1},Z_{j1}X_{j2},Z_{j1}Y_{j2} \dots\}$ (See Fig. \ref{fig:triangular_GSE}). A choice of definition for the Majoranas based on Fenwick trees can give a decrease in the Pauli weight from $O(d)$ to $O(\log(d))$. As such, the Fenwick tree choice will be preferable for graphs of large degree. We refer readers to \cite{setia2019superfast,whitfield2016local} for a discussion of Fenwick tree encodings. The recently proposed ternary tree construction of \cite{jiang2020optimal} would provide a further reduction in Pauli weight.

Each local Majorana is associated to one of the edges incident to the vertex. Each edge then has two local Majoranas associated to it, one from each vertex at its endpoints and we will define our encoded edge operators to be 
\begin{equation}
    \tilde{A}_{jk} = \epsilon_{jk}c_j^{p}c_k^{q}
    \label{eq:encoded_edge}
\end{equation}
where $\epsilon_{jk} = +1$ if $j<k$ and $= -1$ if $j>k$ and $k$ is the $p$-th neighbor of $j$ and $j$ is the $q$-th neighbor of $k$. Encoded vertex operators are a product of all Majoranas at a given vertex,
\begin{equation}
    \tilde{B}_j = i^{d(j)/2} c_j^1 c_j^2 \dots c_j^{d(j)}.
    \label{eq:encoded_vertex}
\end{equation}
If Jordan-Wigner is used to encode the local Majoranas on a given site, then the vertex operator will be Pauli Z acting on all the qubits on that vertex so the occupancy of the mode is stored in the collective parity. Another choice could be made such that the occupancy is stored in a single qubit. One could achieve this an appropriate application of CNOTs.

Finally, for each plaquette $p$ on the graph, we have a stabilizer which is given by a product of all the edge operators around the boundary of the plaquette
\begin{equation}
    S(\partial p) = i^{|\partial p|} \prod_{(j,k)\in \partial p} \tilde{A}_{jk}.
    \label{eq:stabilizer}
\end{equation}
Whereas in the toric code, the plaquette stabilizers detected the presence of a flux, here the plaquette stabilizers detect the presence of a flux without an accompanying charge and vice-versa. In dimensions higher than 1, the logical subspace is that in which charges have fluxes attached. Recall that charge-flux pairs are fermionic in nature which provides the basis for this construction. An example on a triangular lattice is shown in Fig. \ref{fig:triangular_GSE}. For geometries with non-contractible loops, e.g. a torus, we fix boundary conditions to be periodic with a stabilizer consisting of a product of edge operators around the non-contractible loop.

As a final consideration, we consider the odd fermionic parity sector. On a graph with even degree it is not possible to directly encode an odd number of particles. One must introduce an additional non-physical auxiliary mode and create a pair of particles with one occupying the auxiliary mode. The total parity of the physical modes is then odd and the odd-parity simulation can proceed. 
If however there is a vertex on the graph with an odd degree, then there will be a single unpaired Majorana operator on that vertex. Acting with the unpaired Majorana operator can create or destroy a single particle at that vertex without violating any of the stabilizers. As a simple example without any stabilizers, we consider a 1D chain with open boundaries. At the two ends of the chain, there are two unpaired local Majorana operators. If we for example act with the unpaired Majorana on the first site in the chain, we create a single particle. In this way, acting with unpaired Majorana operators changes our parity sector. We can then proceed with our odd-parity sector simulation. In the next section we show that if we pair the local Majorana operators at the ends of the chain, we impose periodic boundary conditions and lose the ability to enter the odd parity sector.

\subsection{1D chain recovers Jordan-Wigner}
To further illustrate the construction in a familiar setting that we hope will give some intuition for the later sections, we will encode a 1D chain of fermions with an onsite potential and periodic boundary conditions. The Hamiltonian of this system is
\begin{align}
    H &= t\sum_j (a^{\dag}_{j+1} a_j + a^{\dag}_j a_{j+1} ) + U\sum_j a^{\dag}_j a_j\\
    &= \frac{-it}{2}\sum_j  (A_{j,j+1}B_{j+1} + B_j A_{j,j+1}) + U\sum_j B_j.
    \label{eq:free_fermions}
\end{align}
Each vertex in this 1D chain obviously has $d=2$ so a single qubit is placed at each vertex. We choose  $c_j^1 = Y_j$, $c_j^2 = X_j$ for the Majoranas at each site. All the edge operators are then $\tilde{A}_{j,j+1} = X_{j}Y_{j+1}$ and the vertex operators are $\tilde{B}_j = Z_j$. The transformed Hamiltonian then takes the familiar form of the XY chain 
\begin{equation}
    \tilde{H} = \frac{t}{2}\sum_j (X_{j}X_{j+1} + Y_{j}Y_{j+1}) + U\sum_j Z_j.
    \label{eq: XY_chain}
\end{equation}
Notice that given the basis chosen for the local Majoranas at each vertex, the Jordan-Wigner transformation is recovered. Indeed, an edge operator between two modes not nearest-neighbor connected is necessarily a product of edge operators in a path connecting the two targeted vertices
\begin{align}
    A_{j,j+n} &= A_{j,j+1}\dots A_{j+n-1,j+n}\\
    \tilde{A}_{j,j+n} &= (-i)^{n-1 }X_j Z_{j+1} \dots Z_{j+n-1} Y_{j+n}
    \label{eq:JW_strings}
\end{align}
giving back the Jordan-Wigner strings of Pauli Zs that all local fermionic encodings are attempting the alleviate.

Finally, as we have periodic boundary conditions and thus a (large) loop in our interaction graph, we have a stabilizer which is the product of all edge operators around the loop. This operator, which is given by $S = \prod_j Z_j$, corresponds to the fact that global fermionic parity is preserved and constrained to be even. We could as well choose not to restrict to the subspace stabilized by the loop as previous described in our discussion of the odd parity sector. In that case, edge operators coupling sites $1,N$ would have a Pauli weight extensive in the system size. The total fermionic parity will always be a symmetry of the Hamiltonian by virtue of only even parity operators being physical. The above shows a consequence of imposing the total parity to be a symmetry of the states as well. This clear physical interpretation does not however generalize to plaquette stabilizers in higher dimensions.

\section{Main result: Custom fermionic codes}

\begin{figure}[h]
    \begin{subfigure}{.48\textwidth}
    \centering
    \includegraphics[width = 0.9\linewidth]{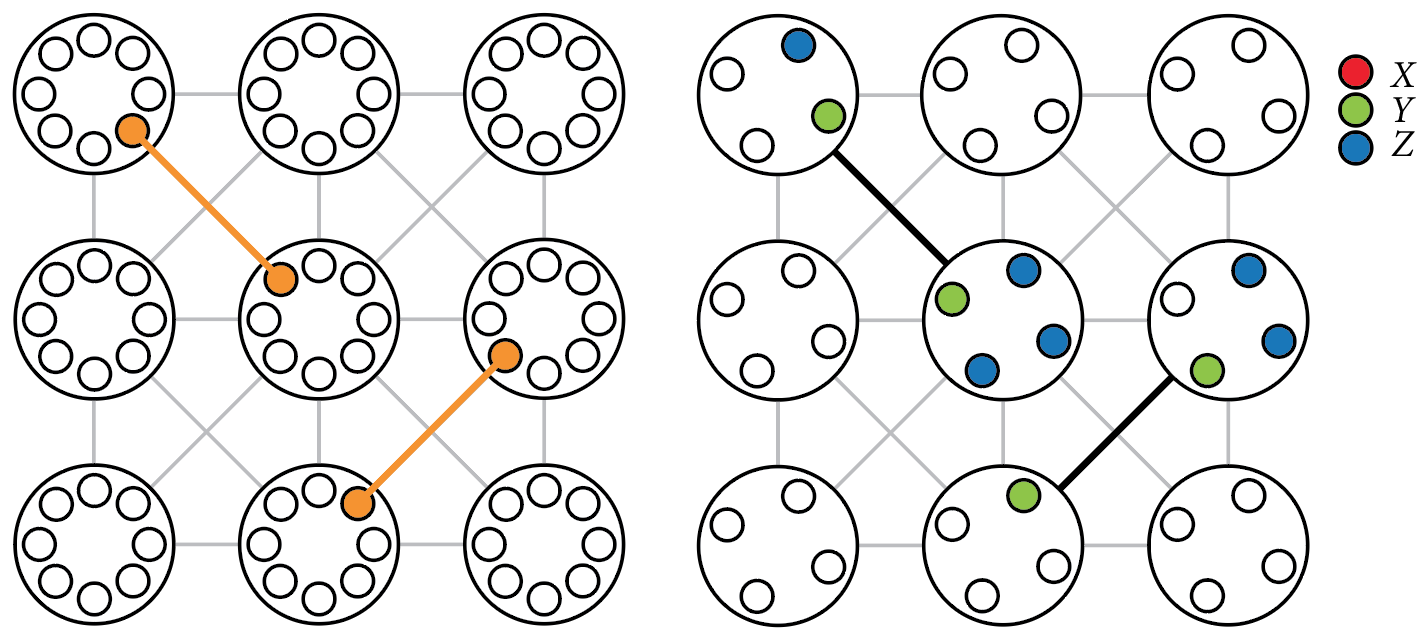}
    \caption{}
    \end{subfigure}
    \begin{subfigure}{0.48\textwidth}
    \centering
    \includegraphics[width = 0.9\linewidth]{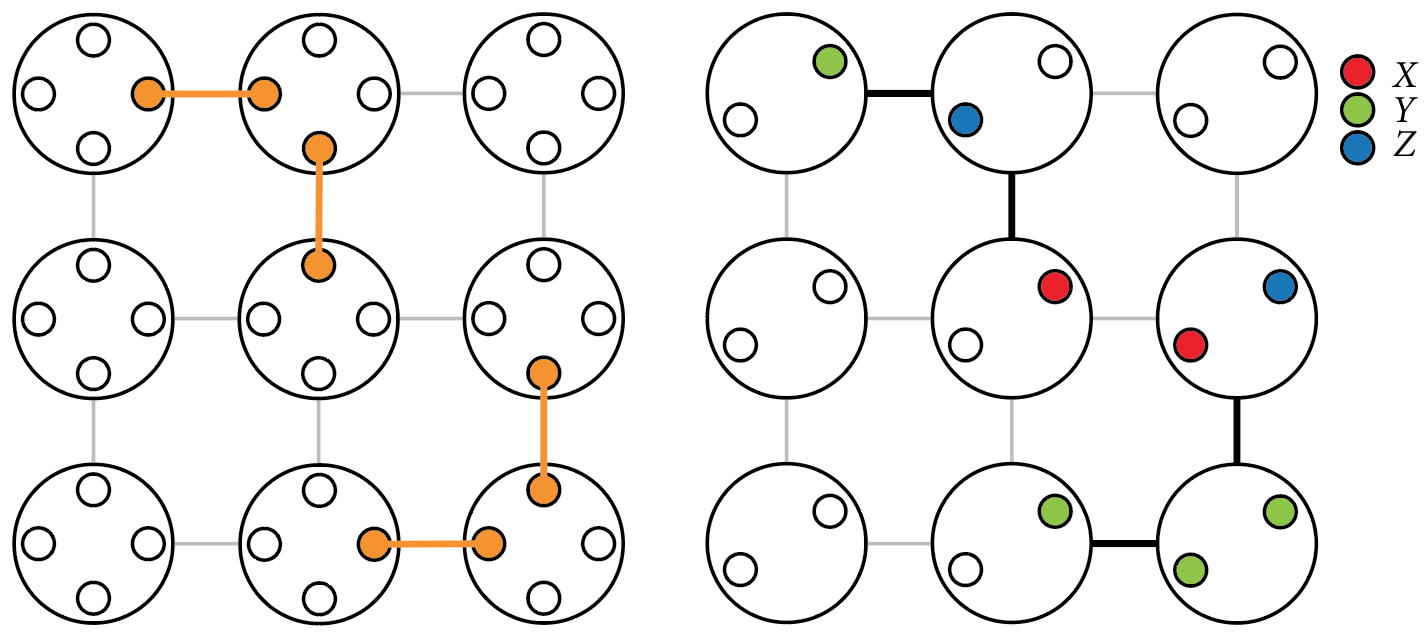}
    \caption{}
    \end{subfigure}
    \caption{(a) Two nearest-diagonal-neighbor couplings (left) fermion picture (right) qubit picture (b) The same two nearest-diagonal-neighbor couplings with diagonal edges omitted from the system graph.}
    \label{fig:next-nearest_edge_omitted}
\end{figure}

Our main result is centered on the idea that regardless of the interaction graph determined by the Hamiltonian terms, we may choose to encode the system into whatever geometry we wish. As such, we will begin discussing two separate graphs for the remainder of the paper, the interaction graph as determined by the Hamiltonian and the system graph that we will encode with our qubit system. The system graph must have at least as many vertices as the interaction graph in order to accommodate the fermionic modes. Also, if an edge operator coupling two modes is present in the Hamiltonian is expressed using the operators of (\ref{eq:edge_op} - \ref{eq:vertex_op}), then a path must connect those two vertices on the system graph. Otherwise the system graph may be arbitrary.

We will discuss a number of useful modifications of the interaction graph. We will briefly discuss sparsifications of the interaction graph - simply omitting edges in the case of lattice models. We then discuss using a virtual geometry including virtual modes that provide paths across which interactions may take place. We there give an example of a system featuring all-to-all interactions.

Next, we give a construction that allows for balancing qubit resource requirements and code locality in the case of lattice models through a blocking construction. We will find it convenient at that time to discuss state preparation. Finally, we will discuss encodings in the presence of constraints on connectivity between qubits using the recently proposed heavy-hexagon lattice as an example.



\subsection{Omitting edges}

As shown above in (\ref{eq:JW_strings}), if two modes are meant to be coupled with an edge operator but the two modes do not share an edge on the system graph, the two modes are still able to be coupled together, but the interaction will be not strictly local. This will lead to a generalization of the JW Z string where the intermediary vertices will all be acted on by a product of two local Majoranas. These generalized Jordan-Wigner strings are similar to string operators in quantum error correcting codes.

For concreteness, consider a $L\times L$ square lattice of fermionic modes interacting with nearest-neighbors and nearest-diagonal-neighbors, those across the diagonal of a square. An example of such a problem is the nearest-diagonal-neighbor Hubbard model. The vertices of the interaction graph are then of degree $d(j)=8$. If we choose the system graph to match the interaction graph, then we require $4$ qubits at every vertex giving $4L^2$ qubits in total. As shown in Fig. \ref{fig:next-nearest_edge_omitted}b, we omit the diagonal edges from our system graph such that each vertex has only degree $d(j)=4$. The diagonal edge operators are then a product of two nearest-neighbor edge operators. The path taken around the square plaquette does not matter as the upper and lower path in each case are equal up to multiplication by a stabilizer. We see that in this case, the Pauli Weight of the qubit operators is smaller for the two paths without the diagonal edges in the system graph. This can be seen in Fig. \ref{fig:next-nearest_edge_omitted}.
Also the presence of the additional qubits also increases the Pauli weight of the nearest-neighbor edge operators and the vertex operators. So, we have shown that strict locality is not always optimal and relaxing to quasi-locality is in some cases beneficial.

The system graph can be sparsified arbitrarily relative to the interaction graph to save qubits as long as paths connect modes that must be coupled with edge operators. In this construction, the qubit requirement is determined by the vertex degrees and not the number of edges as in the local mappings of \cite{bravyi2002fermionic,chen2018exact}. Thus, if a reduction of qubit requirement is sought, one should delete edges with the aim of reducing the degrees of vertices with target degrees being even numbers. 

We would like to mention here that if our system has a square lattice interaction graph and we sparsify the graph in certain ways, we recover the auxiliary qubit mappings of Steudtner and Wehner \cite{steudtnerAQM}. Thus, this construction contains the auxiliary qubit mappings as special case.

\subsection{Adding virtual modes}

\begin{figure}
    \centering
    \includegraphics[width = 0.9\linewidth]{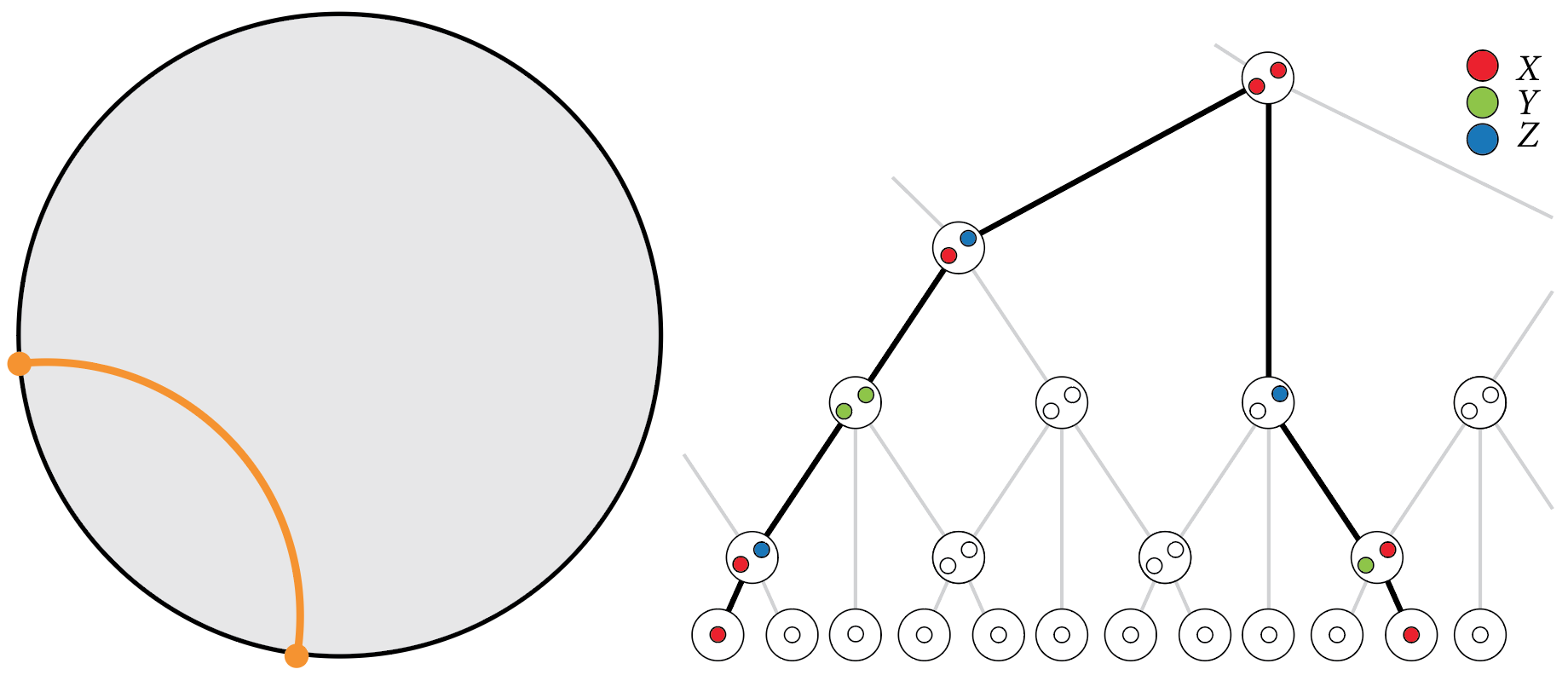}
    \caption{(left) Schematic of an operator coupling two modes at the boundary of the MERA geometry. The orange arc connecting the two endpoints indicates the path of the generalized Jordan-Wigner string. The path through the virtual space depicted in gray is shorter than the path along the boundary. (right) The qubit operator corresponding to a coupling with the generalized Jordan-Wigner string following the discrete geodesic.}
    \label{fig:mera_geodesic}
\end{figure}

Additional virtual modes can be added to the system graph which can in some cases lead to a reduction in the Pauli weight of the transformed Hamiltonians. As always we require closed loops in the system graph to be stabilized by the corresponding plaquette operators. Virtual modes are stabilized by their vertex operators as they will always be unoccupied and so have parity $+1$.

Using virtual modes to reduce Pauli weights could be especially useful in cases where one has a complete or nearly complete interaction graph. Nearly complete interaction graphs (within spin sectors) are known to arise in small molecular Hamiltonians using atomic orbital basis sets leading to large simulation costs with strictly local encodings \cite{chien2019analysis}. A number of all-to-all interacting fermionic models have also become popular in recent years in the study of scrambling of quantum information and of AdS/CFT, the most notable of which is the SYK model \cite{SY_model,Sachdev_blackhole}.

\subsubsection{All-to-all coupled fermions}

We now give an example of a system for which the quantum simulation cost is decreased by using virtual modes. The SYK model consists of $2N$ Majorana fermions with random strength $q$-body interactions coupling all Majoranas. Proposals regarding quantum simulation of the $q=4$ SYK model have previously been put forward \cite{garcia2017digitalSYK,babbush2019quantumSYK,cao2020towardsSYK}.

We will consider the $q=2$ case,
\begin{equation}
    H = -i\sum_{j < k}^{2N} J_{jk}\gamma_{j}\gamma_{k}.
    \label{eq:all-to-all_Hamiltonian}
\end{equation}
We will pair the $2N$ Majorana fermions into $N$ complex fermions and let the indices now run over complex fermions. It will also be convenient for us to break the terms up by whether the Majoranas are odd or even indexed
\begin{multline}
    H =\\ -i \sum_{j < k}^{N} J_{2j-1,2k-1} \godd{j}\godd{k} -i\sum_{j \leq k}^N J_{2j-1,2k} \godd{j}\gev{k}\\
    -i\sum_{j<k}^N J_{2j,2k-1} \gev{j}\godd{k}
    -i \sum_{j < k}^{N} J_{2j,2k} \gev{j}\gev{k}
\end{multline}
We can then write the Hamiltonian in terms of edge and vertex operators
\begin{multline}
    H = \sum_{j<k}^{N} J_{2j-1,2k-1} A_{jk} +\sum_j^N J_{2j-1,2j}B_j\\ +i\sum_{j< k}^N J_{2j-1,2k} A_{jk}B_k
    +i\sum_{j<k}^N J_{2j,2k-1} B_j A_{jk}  \\ -\sum_{j<k}^{N} J_{2j,2k}  A_{jk} B_j B_k 
\end{multline}

The interaction graph is a complete graph so each vertex has degree $d(j)=N-1$. We calculate the qubit requirement and the Pauli Weight for a number of different system geometries):

\begin{enumerate}
    \item A complete graph with $N$ vertices (Complete Graph)
    \item A 1D chain with periodic boundary conditions (Linear)
    \item A geometry with a single virtual mode of degree $N$ connected to all physical modes which only connect to the central virtual mode (N Branches)
    \item Ternary tree with physical modes at the leaves and virtual vertices of degree $4$ (Ternary Tree)
    \item Ternary MERA-like geometry with virtual vertices of degree $4$ (Ternary MERA) (a cutout is shown in Fig. \ref{fig:mera_geodesic})
    \item Hyperbolic geometry where each vertex is degree $6$ and faces are $4$-sided. Physical modes are identified with legs at the boundary of the disk. ($d=6$ Hyperbolic Tiling)
\end{enumerate}

\begin{figure}
    \begin{subfigure}{.48\textwidth}
    \centering
    \includegraphics[width = 0.9\linewidth]{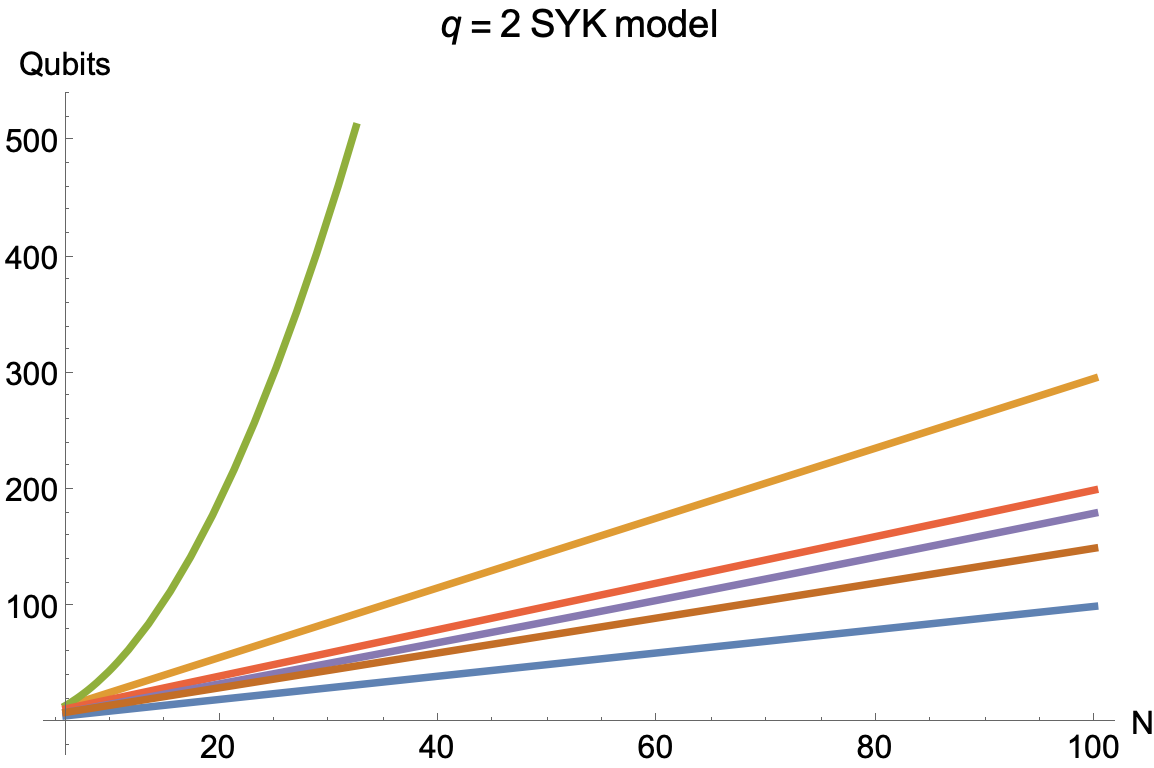}
    \end{subfigure}
    \begin{subfigure}{0.48\textwidth}
    \centering
    \includegraphics[width = 0.94\linewidth]{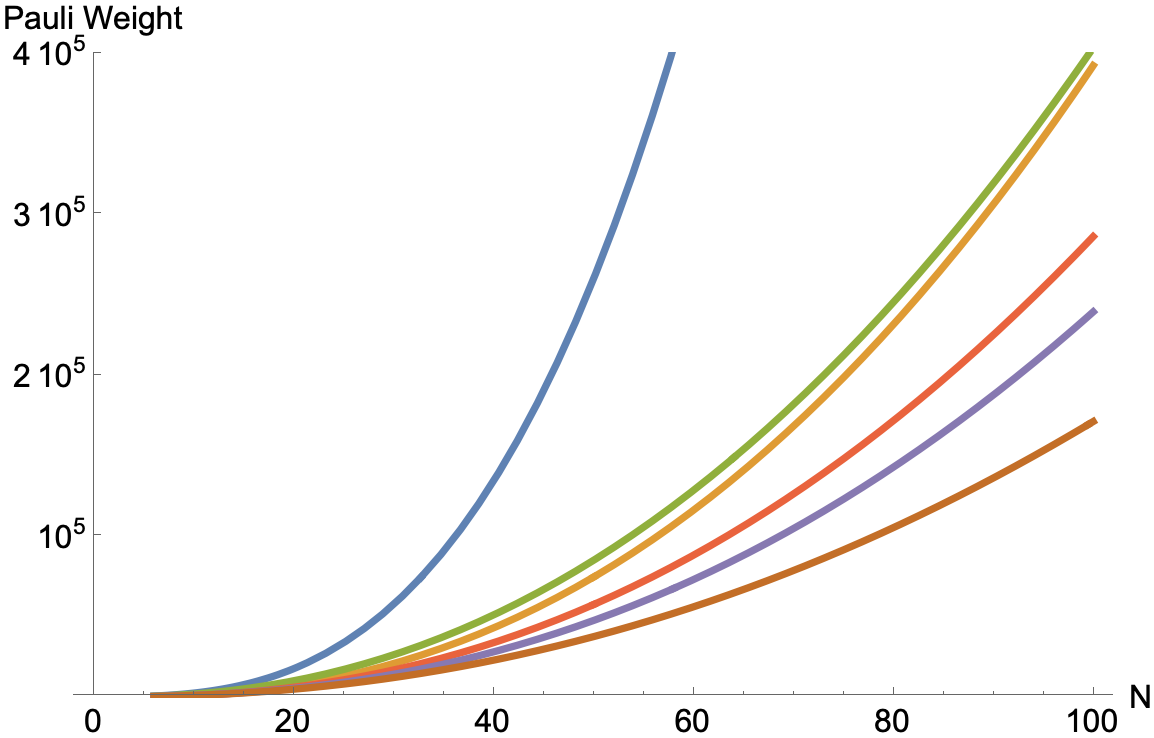}
    \end{subfigure}
    \begin{subfigure}{.48\textwidth}
    \centering
    \includegraphics[width = 0.95
    \linewidth]{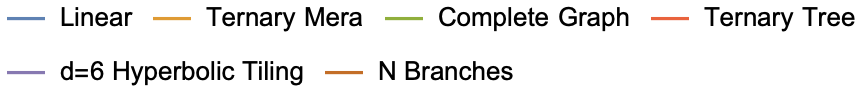}
    \end{subfigure}
    \caption{(Top) Qubit requirements to encode the system of $N$ complex fermions ($2N$ Majorana fermions) given the specified geometry (Bottom) The worst-case Pauli weight of the transformed Hamiltonians.}
    \label{fig:all-to-all_data}
\end{figure}

Some of these geometries have taken their inspiration from tensor networks such as MERA \cite{vidal_1D,swingle2012entanglement} as well as existing literature regarding hyperbolic codes \cite{pastawski2015holographic}. For the geometries containing vertices with vertex degrees growing with system size, we use the Fenwick tree as proposed in \cite{setia2019superfast} which reduces the worst-case weight of operators to logarithmic in the degree. Note also that the ternary tree geometry used here is unrelated to the ternary tree geometry construction of \cite{jiang2020optimal}.

We present the qubit requirements and the worst-case Pauli weight for each geometry vs the number of complex fermions $N$ in Fig. \ref{fig:all-to-all_data}. For the tree, MERA, and hyperbolic geometries, the Pauli weight data presented should be interpreted as approximate given that only certain numbers of modes completely fill levels in the hierarchical constructions.

We now summarize the results shown in Fig. \ref{fig:all-to-all_data}. All geometries except the complete graph required a number of qubits scaling linearly in the number of modes. The complete graph qubit requirements scaled quadratically in the number of modes. While all geometries except the linear chain provided a Pauli weight that scaled as $O(N^2 \log{N})$. The linear geometry had a Pauli weight that scaled as $O(N^3)$.

\begin{figure*}
\centering
    \begin{subfigure}[b]{0.3\textwidth}
        \centering
        \includegraphics[height = 5.7cm]{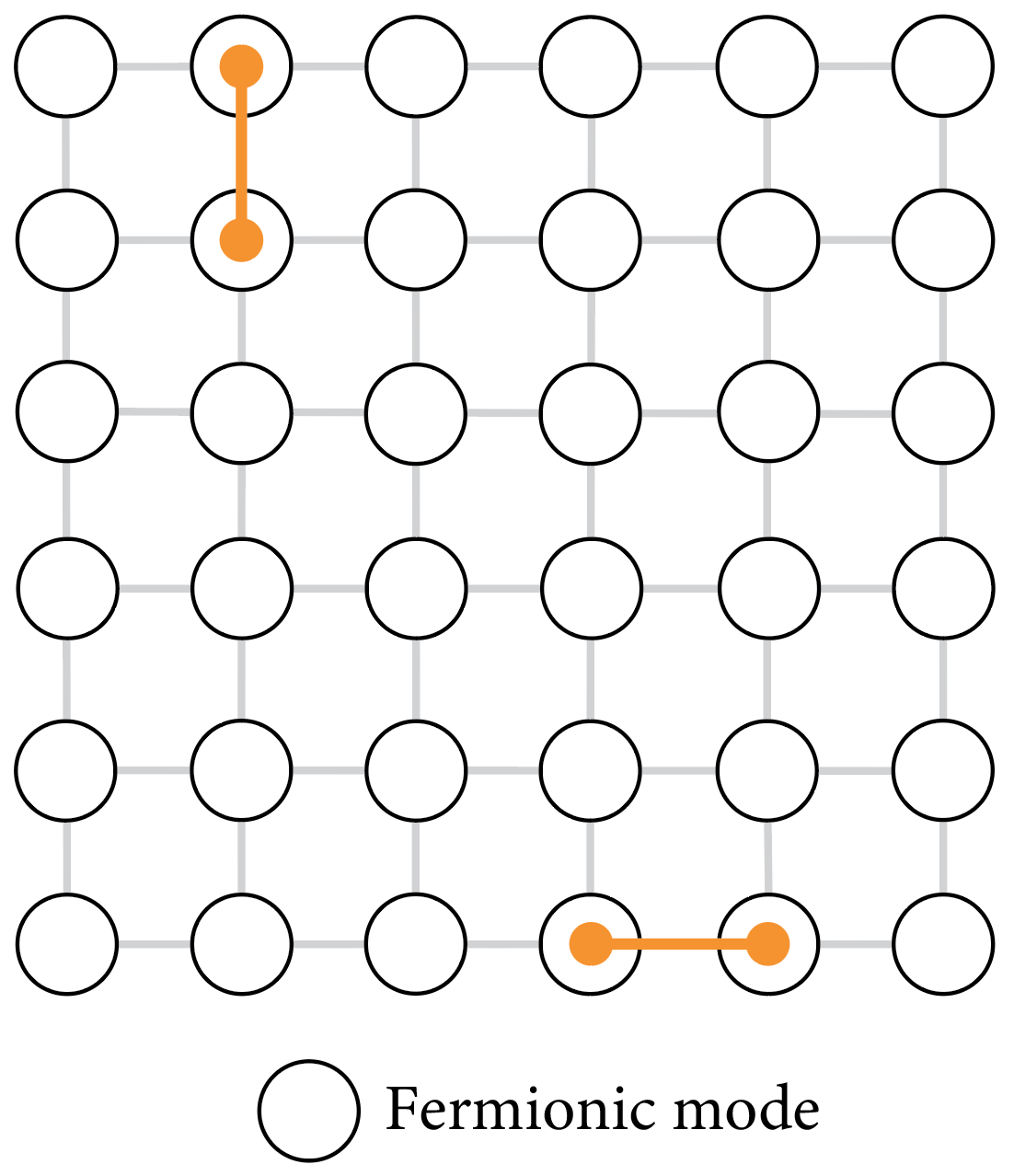}
        \caption{}
    \end{subfigure}%
    ~ 
    \begin{subfigure}[b]{0.3\textwidth}
        \centering
        \includegraphics[height = 5.7cm]{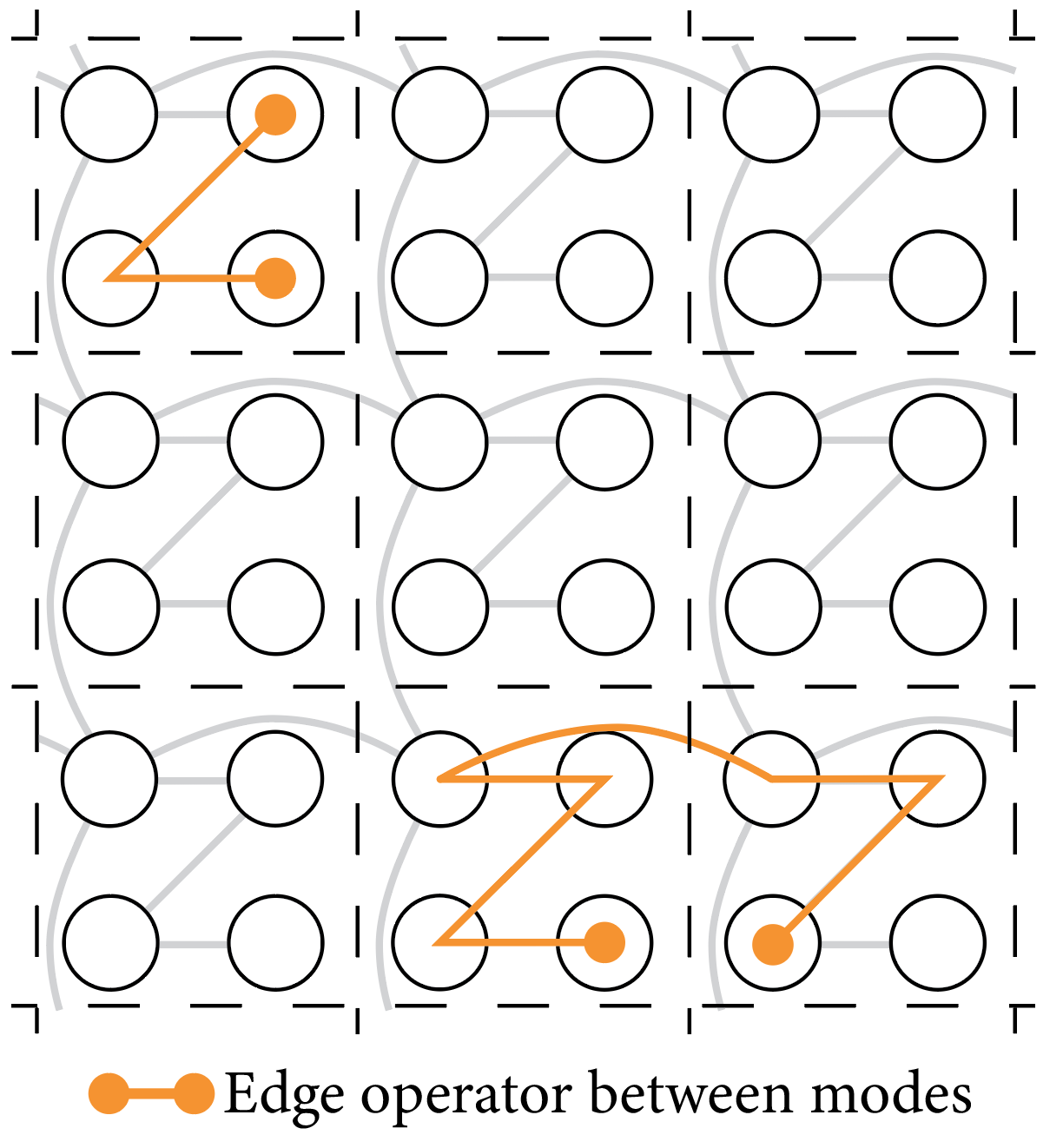}
        \caption{}
    \end{subfigure}
    ~ 
    \begin{subfigure}[b]{0.3\textwidth}
        \centering
        \includegraphics[height = 5.7cm]{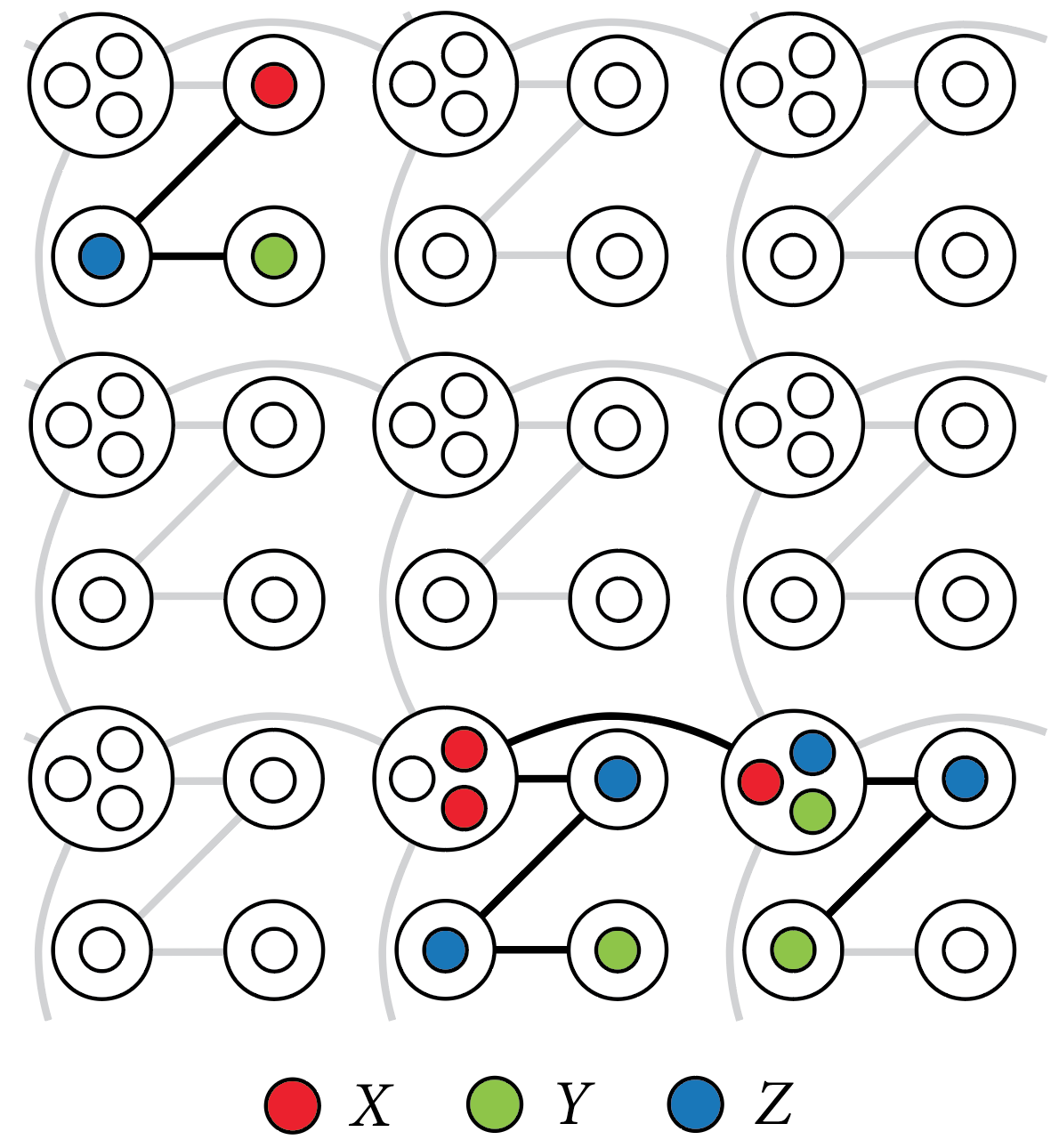}
        \caption{}
    \end{subfigure}
    \caption{(a) A section of a  square lattice of fermionic modes (white circles) with nearest neighbor interaction graph (gray lines). Schematics of two edge operators are shown in orange ( The operators are products of Majoranas from the two fermionic modes).
    (b) The lattice is partitioned into $2\times 2$ blocks of fermionic modes (white). The system graph (gray) is modified from nearest-neighbor square lattice in (a) to a coarser lattice of blocks. The top left mode in each block remains in the lattice while the remaining modes within the block are connected as a 1D chain. The same two edge operators from (a) are shown here. Operators coupling modes within a block run along the chain. Operators coupling sites in adjacent blocks have a generalized Jordan-Wigner string that traverses the lattice from one block to the next. (c) The qubit-encoded version of the system graph is shown. Each fermionic vertex on the lattice, corresponding to fermionic modes, contain either one or three qubits. The qubit operator versions of the edge operators from (a) and (b) are shown. All operators remain local with respect to the coarse-grained lattice.}
    \label{fig: blocked_lattice}
\end{figure*}

In this case, we see that the matching the system graph to the interaction graph exactly is not the most economical encoding strategy. It requires the greatest qubit requirements of all geometries shown, $N(N-1)/2$, while giving a Pauli weight scaling that was comparable to other geometries. The opposite limit, the 1D geometry, requires the least number of qubits, $N$, but due to the non-locality of the string operators, the Pauli weight scales as $O(N^3)$. 

Geometries where we include virtual modes which provide shorter paths between vertices perform favorably at the cost of more qubits. For example, we can include a single central mode connected to all others which serves as a midpoint for interactions. Using this geometry, all couplings are products of two edge operators from the physical modes to the virtual one and we require $1.5N$ qubits. The Pauli weight still scales as $O(N^2 \log{N})$ but with the Fenwick tree encoding of local Majorana operators was the lowest of all geometries tested.

The other three geometries presented in Fig. \ref{fig:all-to-all_data} also all require a number of qubits that is linear in $N$. The hierarchical structure of the geometries means that points separated by $l$ on the boundary have paths through the virtual space that are length $\sim\log{l}$. As a result, they offer Pauli weights that scale as $O(N^2 \log{N})$. In the case of the ternary MERA and $d=6$ hyperbolic geometries we can give a nice interpretation, the couplings feature generalized JW string operators traversing discrete versions of geodesics in the virtual hyperbolic geometry, Fig \ref{fig:mera_geodesic}.

We note that in a number of the geometries investigated above, we placed physical modes at the boundaries of a hyperbolic disk or at the leaves of a tree while the other modes were considered virtual i.e. not corresponding to physical modes. We could also have chosen to associate the vertices in the bulk of graph with physical modes. This would provide a savings in the required number of qubits.

Previous proposals for simulation of the SYK model have used Jordan-Wigner which in the $q=2$ case leads to a Pauli weight of $O(N^3)$. As we have shown this can be reduced to $O(N^2 \log{N})$. Thus we can reduce the complexity of the simulation with a more careful consideration of encoding geometry.

Given that simulations of all-to-all interacting systems can benefit from a virtual geometry, it would be interesting to extend our investigation to sparsified graphs for example the sparsified SYK model proposed in \cite{xu2020sparseSYK}. In addition to contributing towards progress in studies quantum chaos and holography, we also believe that investigations of quantum simulations of such highly connected systems will also benefit simulations of quantum chemistry as chemical Hamiltonians can also feature regions of high connectivity as we have shown in \cite{chien2019analysis}.

\subsection{Qubit vs. locality trade-off with blocking}

We will now consider an approach to finding a balance between locality and qubit requirement for a system on a $L\times L$ square lattice. A linear (JW) geometry will require $L^2$ qubits but will suffer from long JW strings whereas a strictly local encoding will require roughly $2L^2$ qubits (minus a few on the boundaries). Again, we can reduce the required number of qubits by relaxing the necessity for strictly local interactions. We will divide the system into a number of blocks. Interactions within blocks will be non-local incurring Jordan-Wigner strings that increase in length with the size of the blocks. Interactions between blocks, however, remain local in that the Pauli weight of operators are independent of the size of the full lattice. 

The blocking goes as follows. Partition the lattice into the desired number of blocks, $b$, determined by the available resources. Treat the modes within the block as though they were on a 1D chain. The lattice is then a coarser lattice with each vertex connected to a 1D chain. The first mode in the chain remains connected to the lattice such that those vertices are of degree $d=5$ in the bulk of the system and so require $3$ qubits. The total number of qubits required is then $L^2 + 2b$ (up to boundaries). With this construction, we are free to interpolate between a strictly local encoding with $L^2$ blocks of size $1$ and $1$ block of size $L^2$ by choosing blocks to be of the desired size. 

The idea of using a segmented encoding was explored in \cite{whitfield2016local} where segmented versions of the Bravyi-Kitaev and Fenwick tree encodings were explored for the 2D Hubbard model.

\subsubsection{As truncated state preparation}

We now hope to provide an intuitive picture for the blocking construction. This will also be a convenient time to address state preparation. For simplicity, we will again consider a square lattice of dimension $L\times L$. As previously mentioned, this encoding manages to encode the fermions in a local way by representing them as excitations of the toric code which are odd under particle exchange. Thus, to use this encoding, one must prepare a toric code state which is well known to be topologically ordered and therefore long-range entangled \cite{chen2010local}.

Utilizing a MERA quantum circuit, one can prepare a toric code state by introducing entanglement scale-by-scale beginning with long-range entanglement and ending with entanglement at the final lattice scale \cite{aguado2008entanglement}. The circuit is comprised of a number of levels $U = U_1\dots U_{\log{L}}$. Each level $k$ takes as input a state on a lattice of linear size $l$ and a number of ancilla qubits and outputs a state of linear size $2l$,
\begin{equation}
    U_{k} \ket{\psi_l}\ket{0\dots 0} = \ket{\psi_{2l}}.
    \label{eq:mera_circ}
\end{equation}
The state at each level has four times as many plaquettes as the previous level and so, with corrections at the boundaries, has about four times as many qubits in addition to the $L^2$ qubits associated to the fermionic modes. Again, at the final level, the total number of qubits is $2L^2$ with correction at the boundaries.
Each layer consists of acting with Hadamards and CNOTs locally with respect to the lattice at the given level. The exact form of the circuit can be found in \cite{aguado2008entanglement}. Upon application of the $\log{L}$ layers of the circuit, the toric state is prepared. At this point, a constant depth unitary is performed to satisfy the stabilizers which differ slightly from the true toric code model. The exact form of this circuit depends on the basis chosen for the local Majoranas. If a Jordan-Wigner basis is chosen, the circuit consists merely of a single layer of Pauli X and Z gates.

\begin{figure}
    \centering
    \includegraphics[width = 0.75\linewidth]{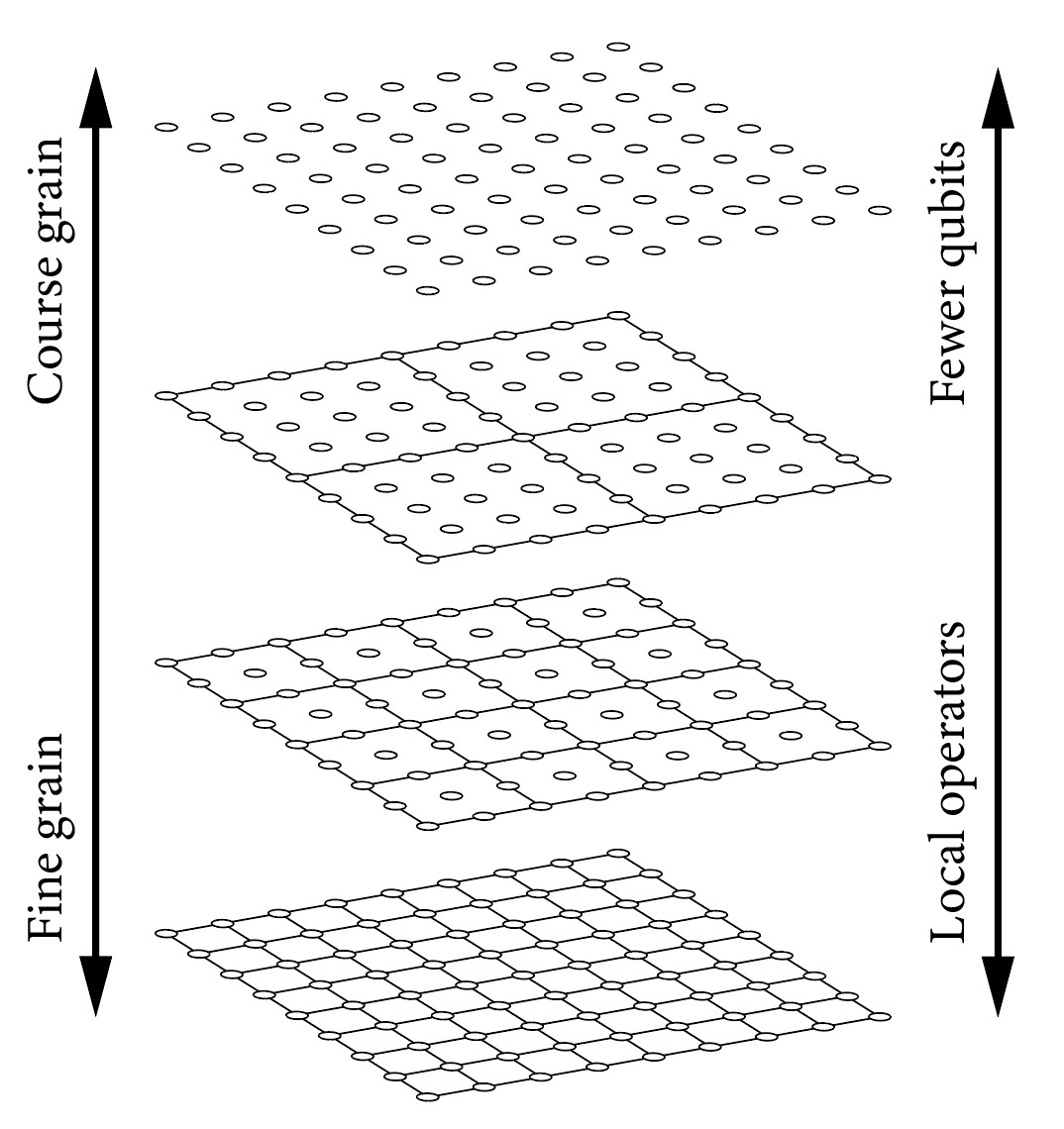}
    \caption{Here we show the trade-off between qubit requirement and operator locality. The circles represent the fermionic modes while the lines represent the lattice of the topologically ordered state underpinning the local encoding. Each level represents a case in which one additional level of the state preparation circuit is applied, creating a finer lattice for the underlying toric code. Operators are local only with respect to the lattice spacing of the toric code state. At the top level with no topological order, the fermionic operators may be fully non-local e.g. long JW strings. At the bottom level, the operators are fully local with respect to the lattice spacing of the fermions but twice as many qubits are required. }
    \label{fig:coarse_grain}
\end{figure}

The blocking construction can be thought of as a truncation of the state preparation circuit. Truncating the circuit results in a coarse-grained toric code state relative to the lattice of fermionic modes. As such, the fermionic operators which are local due to the topological order of the toric code state, are now only local with respect to the toric code lattice. Operators may be non-local up to the scale of the toric code lattice spacing. The benefit, however, is a savings in qubit resources as each MERA layer requires additional qubits. Thus by utilizing topological order on a coarse-grained lattice, one can realize the trade-off between operator locality and qubit requirement as depicted in Fig. \ref{fig:coarse_grain}.

Although each level of the MERA unitary is local with respect to the lattice at each level, it is not local with respect to the final lattice of qubits. With strictly local operations, preparing the topologically ordered toric code state takes a time proportional to the linear size of the lattice \cite{bravyi2006lieb}.

\subsubsection{Further generalizations}

We now discuss a number of ways this above construction can be generalized.
Going beyond a square lattice of blocks, one could perform the partitioning of the system of $N$ modes into a set of general sites $S=\{s_1,\dots,s_{|S|}\}$ each containing a number of modes $n(s_i)$ and where each site is connected to $d(s_i)$ others on the lattice of sites. Then given the construction above, the total number of qubits required would be
\begin{equation}
    \# \text{ of qubits } = N + \sum_i^{|S|} \left\lceil\frac{d(s_i)}{2}\right\rceil.
    \label{eq: qubits_general_sites}
\end{equation}

Further, one is free to choose any encoding for the modes within each site. For example, one could choose to encode the modes on a 1D chain as described or choose to use a Fenwick tree \cite{whitfield2016local} or Jiang et al.'s ternary tree encoding \cite{jiang2020optimal}.

To emphasize the generality of the construction we are proposing here and to illustrate how the geometry of the interactions should inform the geometry of the qubit system, we would like to sketch how one might encode a system of interest lately, that being a lattice of SYK islands \cite{gu2017latticeSYK,song2017latticeSYK}. The system is a lattice of islands $S=\{s_1,\dots,s_{|S|}\}$ each containing $n(s_i)$ Majorana fermions with two types of interactions, quartic interactions between Majoranas within each island with random strength and quadratic interactions between Majoranas on adjacent islands on the lattice. We propose that such a system would be best simulated as a lattice of sites, where the modes on each site, are placed at vertices on the ``$N$ branches'' geometry or a hierarchical geometry as described above. The chosen geometry is then attached to the lattice at the central vertex. We highlight such a system as it frustrates many of the existing encoding schemes, featuring both highly connected regions as well as a notion of locality.

Finally, we reiterate that one is free to use any basis for the local Majorana operators at each vertex. A Jordan-Wigner encoding was chosen for simplicity but an improvement in Pauli weight can be achieved by using a different local basis. For example, a Fenwick tree basis for local Majorana operators as proposed in \cite{setia2019superfast} or Jiang et al.'s ternary tree basis \cite{jiang2020optimal} would give a Pauli weight for Majorana operators logarithmic in the number of qubits.

\subsection{Device connectivity constraints}

\begin{figure}
    \centering
    \includegraphics[width = 0.9\linewidth]{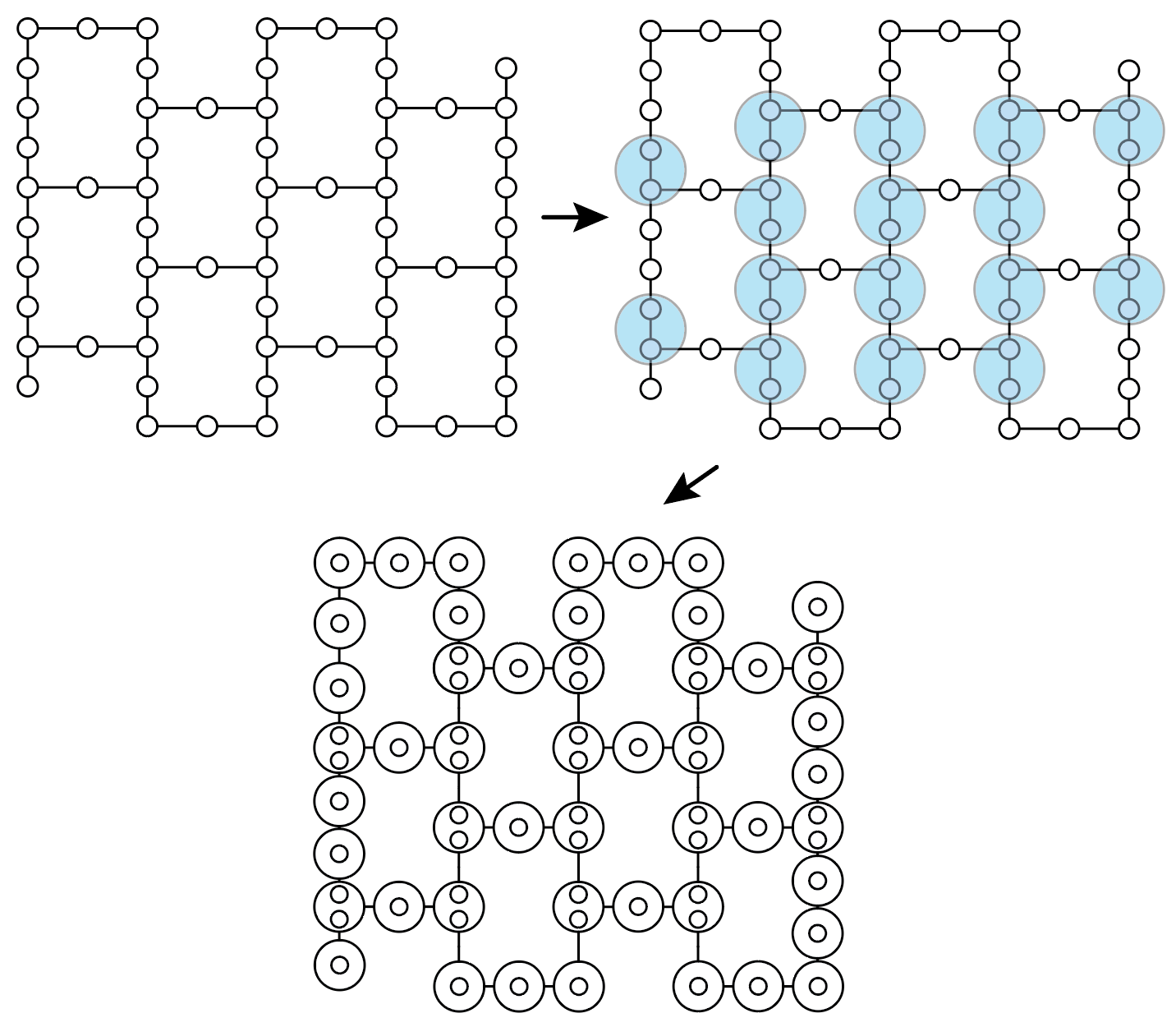}
    \caption{(top left) The heavy-hexagon geometry as presented in \cite{chamberland2020topological}. 65 qubits are shown. (top right) We show a choice of grouping pairs of qubits together to encode the degree $3$ vertices. (bottom) We show the system graph where each larger circle corresponds to one of the 49 encoded fermionic modes. Smaller circles represent the qubits associated to each mode.}
    \label{fig:heavy_hex}
\end{figure}

Many of the quantum computing platforms under development are subject to connectivity constraints. Notably, these include superconducting qubits which have recently been used to achieve a ``quantum supremacy'' result \cite{arute2019quantumsupremacy}. The processor used in the supremacy experiment features qubits laid out on a square grid with nearest-neighbor connectivity. Recent work on increasing the capabilities of quantum computers as measured in so-called quantum volume \cite{cross2019validating} has led to progress on devices with qubits laid out on lower-degree graphs. In particular, the heavy-hexagon lattice has been identified as a candidate system geometry for realizing quantum error correction while mitigating hardware challenges presented by cross-talk and frequency collisions \cite{chamberland2020topological}. This lattice features qubits placed on vertices of a hexagonal lattice as well as on edges.

On such devices subject to connectivity constraints, it may be preferable to encode a fermionic system into a graph informed by the connectivity of the device. To that end, we present here a candidate geometry for encoding a fermionic system into a heavy-hexagon lattice. In Fig. \ref{fig:heavy_hex}, we show a $65$ qubit heavy-hexagon lattice. To each degree $3$ vertex, we associate an additional qubit. Thus, with $16$ degree $3$ vertices, we are able to encode 49 fermionic modes into the 65 qubit heavy-hexagon lattice. As above, coupling modes which do not share an edge in the system geometry will require edge operators containing generalized Jordan-Wigner strings. 

Designing device-specific encodings for other platforms would proceed similarly, identifying qubits to group together to encode vertices of appropriate degree and embedding a problem withing the device-informed geometry. We leave an investigation of optimal device-specific geometries to future work.

\section{Conclusion}

In this paper, we have presented a very general construction for designing quantum codes to simulate fermions on quantum computers. The construction realizes the trade-off between qubit resources and operator locality in such a way that one can tailor the encoding to best fit the resources at hand. We have also shown that in some cases, locality is too strict of a constraint and one is better off seeking a quasi-local encoding. We showed this occurs in systems such as square lattice models with nearest- and nearest-diagonal-neighbor coupling where it is best to simply encode the square lattice and realize the diagonal couplings quasi-locally. We also presented the case of a fermionic system with all-to-all connectivity and demonstrated that one should encode this system with a virtual geometry so that generalized Jordan-Wigner strings traverse paths through the virtual geometry. We discussed how quasi-local codes can be interpreted as local codes with truncated state preparation circuits. Finally, we considered the design of custom codes to suit device connectivity constraints. We expect that the encoding construction presented here will find use in quantum simulation experiments ranging from quantum chemistry, to quantum gravity, to condensed matter physics.

\section*{Acknowledgements}

RWC and JDW were funded by the NSF (PHYS-1820747) and the Department of Energy (Grant DE-SC0019374). JDW is also supported by NSF (EPSCoR-1921199) and by the U.S. Department of Energy, Office of Science, Office of Advanced Scientific Computing Research under programs Quantum Computing Application Teams and Accelerated Research for Quantum Computing program.

\bibliographystyle{apsrev}
\bibliography{main}

\begin{thebibliography}{47}
\expandafter\ifx\csname natexlab\endcsname\relax\def\natexlab#1{#1}\fi
\expandafter\ifx\csname bibnamefont\endcsname\relax
  \def\bibnamefont#1{#1}\fi
\expandafter\ifx\csname bibfnamefont\endcsname\relax
  \def\bibfnamefont#1{#1}\fi
\expandafter\ifx\csname citenamefont\endcsname\relax
  \def\citenamefont#1{#1}\fi
\expandafter\ifx\csname url\endcsname\relax
  \def\url#1{\texttt{#1}}\fi
\expandafter\ifx\csname urlprefix\endcsname\relax\def\urlprefix{URL }\fi
\providecommand{\bibinfo}[2]{#2}
\providecommand{\eprint}[2][]{\url{#2}}

\bibitem[{\citenamefont{Wigner and Jordan}(1928)}]{jordanwigner}
\bibinfo{author}{\bibfnamefont{E.~P.} \bibnamefont{Wigner}} \bibnamefont{and}
  \bibinfo{author}{\bibfnamefont{P.}~\bibnamefont{Jordan}},
  \bibinfo{journal}{Z. Phys} \textbf{\bibinfo{volume}{47}},
  \bibinfo{pages}{631} (\bibinfo{year}{1928}).

\bibitem[{\citenamefont{Lieb et~al.}(1961)\citenamefont{Lieb, Schultz, and
  Mattis}}]{lieb1961two}
\bibinfo{author}{\bibfnamefont{E.}~\bibnamefont{Lieb}},
  \bibinfo{author}{\bibfnamefont{T.}~\bibnamefont{Schultz}}, \bibnamefont{and}
  \bibinfo{author}{\bibfnamefont{D.}~\bibnamefont{Mattis}},
  \bibinfo{journal}{Annals of Physics} \textbf{\bibinfo{volume}{16}},
  \bibinfo{pages}{407} (\bibinfo{year}{1961}).

\bibitem[{\citenamefont{Feynman}(1999)}]{feynman1999simulating}
\bibinfo{author}{\bibfnamefont{R.~P.} \bibnamefont{Feynman}},
  \bibinfo{journal}{Int. J. Theor. Phys} \textbf{\bibinfo{volume}{21}}
  (\bibinfo{year}{1999}).

\bibitem[{\citenamefont{Lloyd}(1996)}]{lloyd1996universal}
\bibinfo{author}{\bibfnamefont{S.}~\bibnamefont{Lloyd}},
  \bibinfo{journal}{Science} pp. \bibinfo{pages}{1073--1078}
  (\bibinfo{year}{1996}).

\bibitem[{\citenamefont{McArdle et~al.}(2020)\citenamefont{McArdle, Endo,
  Aspuru-Guzik, Benjamin, and Yuan}}]{mcardle2020quantum}
\bibinfo{author}{\bibfnamefont{S.}~\bibnamefont{McArdle}},
  \bibinfo{author}{\bibfnamefont{S.}~\bibnamefont{Endo}},
  \bibinfo{author}{\bibfnamefont{A.}~\bibnamefont{Aspuru-Guzik}},
  \bibinfo{author}{\bibfnamefont{S.~C.} \bibnamefont{Benjamin}},
  \bibnamefont{and} \bibinfo{author}{\bibfnamefont{X.}~\bibnamefont{Yuan}},
  \bibinfo{journal}{Reviews of Modern Physics} \textbf{\bibinfo{volume}{92}},
  \bibinfo{pages}{015003} (\bibinfo{year}{2020}).

\bibitem[{\citenamefont{Reiher et~al.}(2017)\citenamefont{Reiher, Wiebe, Svore,
  Wecker, and Troyer}}]{reiher2017elucidating}
\bibinfo{author}{\bibfnamefont{M.}~\bibnamefont{Reiher}},
  \bibinfo{author}{\bibfnamefont{N.}~\bibnamefont{Wiebe}},
  \bibinfo{author}{\bibfnamefont{K.~M.} \bibnamefont{Svore}},
  \bibinfo{author}{\bibfnamefont{D.}~\bibnamefont{Wecker}}, \bibnamefont{and}
  \bibinfo{author}{\bibfnamefont{M.}~\bibnamefont{Troyer}},
  \bibinfo{journal}{Proceedings of the National Academy of Sciences}
  \textbf{\bibinfo{volume}{114}}, \bibinfo{pages}{7555} (\bibinfo{year}{2017}).

\bibitem[{\citenamefont{Abrams and Lloyd}(1997)}]{abrams1997simulation}
\bibinfo{author}{\bibfnamefont{D.~S.} \bibnamefont{Abrams}} \bibnamefont{and}
  \bibinfo{author}{\bibfnamefont{S.}~\bibnamefont{Lloyd}},
  \bibinfo{journal}{Physical Review Letters} \textbf{\bibinfo{volume}{79}},
  \bibinfo{pages}{2586} (\bibinfo{year}{1997}).

\bibitem[{\citenamefont{Motta et~al.}(2020)\citenamefont{Motta, Sun, Tan,
  O’Rourke, Ye, Minnich, Brand{\~a}o, and Chan}}]{motta2020imaginary}
\bibinfo{author}{\bibfnamefont{M.}~\bibnamefont{Motta}},
  \bibinfo{author}{\bibfnamefont{C.}~\bibnamefont{Sun}},
  \bibinfo{author}{\bibfnamefont{A.~T.} \bibnamefont{Tan}},
  \bibinfo{author}{\bibfnamefont{M.~J.} \bibnamefont{O’Rourke}},
  \bibinfo{author}{\bibfnamefont{E.}~\bibnamefont{Ye}},
  \bibinfo{author}{\bibfnamefont{A.~J.} \bibnamefont{Minnich}},
  \bibinfo{author}{\bibfnamefont{F.~G.} \bibnamefont{Brand{\~a}o}},
  \bibnamefont{and} \bibinfo{author}{\bibfnamefont{G.~K.-L.}
  \bibnamefont{Chan}}, \bibinfo{journal}{Nature Physics}
  \textbf{\bibinfo{volume}{16}}, \bibinfo{pages}{205} (\bibinfo{year}{2020}).

\bibitem[{\citenamefont{Bravyi and Kitaev}(2005)}]{bravyi2005universal}
\bibinfo{author}{\bibfnamefont{S.}~\bibnamefont{Bravyi}} \bibnamefont{and}
  \bibinfo{author}{\bibfnamefont{A.}~\bibnamefont{Kitaev}},
  \bibinfo{journal}{Physical Review A} \textbf{\bibinfo{volume}{71}},
  \bibinfo{pages}{022316} (\bibinfo{year}{2005}).

\bibitem[{\citenamefont{Bravyi and Kitaev}(2002)}]{bravyi2002fermionic}
\bibinfo{author}{\bibfnamefont{S.~B.} \bibnamefont{Bravyi}} \bibnamefont{and}
  \bibinfo{author}{\bibfnamefont{A.~Y.} \bibnamefont{Kitaev}},
  \bibinfo{journal}{Annals of Physics} \textbf{\bibinfo{volume}{298}},
  \bibinfo{pages}{210} (\bibinfo{year}{2002}).

\bibitem[{\citenamefont{Jiang et~al.}(2020)\citenamefont{Jiang, Kalev,
  Mruczkiewicz, and Neven}}]{jiang2020optimal}
\bibinfo{author}{\bibfnamefont{Z.}~\bibnamefont{Jiang}},
  \bibinfo{author}{\bibfnamefont{A.}~\bibnamefont{Kalev}},
  \bibinfo{author}{\bibfnamefont{W.}~\bibnamefont{Mruczkiewicz}},
  \bibnamefont{and} \bibinfo{author}{\bibfnamefont{H.}~\bibnamefont{Neven}},
  \bibinfo{journal}{Quantum} \textbf{\bibinfo{volume}{4}}, \bibinfo{pages}{276}
  (\bibinfo{year}{2020}).

\bibitem[{\citenamefont{Ball}(2005)}]{ball2005fermions}
\bibinfo{author}{\bibfnamefont{R.}~\bibnamefont{Ball}},
  \bibinfo{journal}{Physical review letters} \textbf{\bibinfo{volume}{95}},
  \bibinfo{pages}{176407} (\bibinfo{year}{2005}).

\bibitem[{\citenamefont{Verstraete and Cirac}(2005)}]{verstraete2005mapping}
\bibinfo{author}{\bibfnamefont{F.}~\bibnamefont{Verstraete}} \bibnamefont{and}
  \bibinfo{author}{\bibfnamefont{J.~I.} \bibnamefont{Cirac}},
  \bibinfo{journal}{Journal of Statistical Mechanics: Theory and Experiment}
  \textbf{\bibinfo{volume}{2005}}, \bibinfo{pages}{P09012}
  (\bibinfo{year}{2005}).

\bibitem[{\citenamefont{Whitfield et~al.}(2016)\citenamefont{Whitfield,
  Havl{\'\i}{\v{c}}ek, and Troyer}}]{whitfield2016local}
\bibinfo{author}{\bibfnamefont{J.~D.} \bibnamefont{Whitfield}},
  \bibinfo{author}{\bibfnamefont{V.}~\bibnamefont{Havl{\'\i}{\v{c}}ek}},
  \bibnamefont{and} \bibinfo{author}{\bibfnamefont{M.}~\bibnamefont{Troyer}},
  \bibinfo{journal}{Physical Review A} \textbf{\bibinfo{volume}{94}},
  \bibinfo{pages}{030301} (\bibinfo{year}{2016}).

\bibitem[{\citenamefont{Setia and Whitfield}(2018)}]{setia2018}
\bibinfo{author}{\bibfnamefont{K.}~\bibnamefont{Setia}} \bibnamefont{and}
  \bibinfo{author}{\bibfnamefont{J.~D.} \bibnamefont{Whitfield}},
  \bibinfo{journal}{The Journal of Chemical Physics}
  \textbf{\bibinfo{volume}{148}}, \bibinfo{pages}{164104}
  (\bibinfo{year}{2018}).

\bibitem[{\citenamefont{Chien et~al.}(2019)\citenamefont{Chien, Xue, Hardikar,
  Setia, and Whitfield}}]{chien2019analysis}
\bibinfo{author}{\bibfnamefont{R.~W.} \bibnamefont{Chien}},
  \bibinfo{author}{\bibfnamefont{S.}~\bibnamefont{Xue}},
  \bibinfo{author}{\bibfnamefont{T.~S.} \bibnamefont{Hardikar}},
  \bibinfo{author}{\bibfnamefont{K.}~\bibnamefont{Setia}}, \bibnamefont{and}
  \bibinfo{author}{\bibfnamefont{J.~D.} \bibnamefont{Whitfield}},
  \bibinfo{journal}{Physical Review A} \textbf{\bibinfo{volume}{100}},
  \bibinfo{pages}{032337} (\bibinfo{year}{2019}).

\bibitem[{\citenamefont{Setia et~al.}(2019)\citenamefont{Setia, Bravyi,
  Mezzacapo, and Whitfield}}]{setia2019superfast}
\bibinfo{author}{\bibfnamefont{K.}~\bibnamefont{Setia}},
  \bibinfo{author}{\bibfnamefont{S.}~\bibnamefont{Bravyi}},
  \bibinfo{author}{\bibfnamefont{A.}~\bibnamefont{Mezzacapo}},
  \bibnamefont{and} \bibinfo{author}{\bibfnamefont{J.~D.}
  \bibnamefont{Whitfield}}, \bibinfo{journal}{Physical Review Research}
  \textbf{\bibinfo{volume}{1}}, \bibinfo{pages}{033033} (\bibinfo{year}{2019}).

\bibitem[{\citenamefont{Jiang et~al.}(2019)\citenamefont{Jiang, McClean,
  Babbush, and Neven}}]{jiang2019majorana}
\bibinfo{author}{\bibfnamefont{Z.}~\bibnamefont{Jiang}},
  \bibinfo{author}{\bibfnamefont{J.}~\bibnamefont{McClean}},
  \bibinfo{author}{\bibfnamefont{R.}~\bibnamefont{Babbush}}, \bibnamefont{and}
  \bibinfo{author}{\bibfnamefont{H.}~\bibnamefont{Neven}},
  \bibinfo{journal}{Physical Review Applied} \textbf{\bibinfo{volume}{12}},
  \bibinfo{pages}{064041} (\bibinfo{year}{2019}).

\bibitem[{\citenamefont{Steudtner and Wehner}(2019)}]{steudtnerAQM}
\bibinfo{author}{\bibfnamefont{M.}~\bibnamefont{Steudtner}} \bibnamefont{and}
  \bibinfo{author}{\bibfnamefont{S.}~\bibnamefont{Wehner}},
  \bibinfo{journal}{Physical Review A} \textbf{\bibinfo{volume}{99}},
  \bibinfo{pages}{022308} (\bibinfo{year}{2019}).

\bibitem[{\citenamefont{Derby and Klassen}(2020)}]{derby2020low}
\bibinfo{author}{\bibfnamefont{C.}~\bibnamefont{Derby}} \bibnamefont{and}
  \bibinfo{author}{\bibfnamefont{J.}~\bibnamefont{Klassen}},
  \bibinfo{journal}{arXiv preprint arXiv:2003.06939}  (\bibinfo{year}{2020}).

\bibitem[{\citenamefont{Chen et~al.}(2018)\citenamefont{Chen, Kapustin, and
  Radi{\v{c}}evi{\'c}}}]{chen2018exact}
\bibinfo{author}{\bibfnamefont{Y.-A.} \bibnamefont{Chen}},
  \bibinfo{author}{\bibfnamefont{A.}~\bibnamefont{Kapustin}}, \bibnamefont{and}
  \bibinfo{author}{\bibfnamefont{{\DJ}.}~\bibnamefont{Radi{\v{c}}evi{\'c}}},
  \bibinfo{journal}{Annals of Physics} \textbf{\bibinfo{volume}{393}},
  \bibinfo{pages}{234} (\bibinfo{year}{2018}).

\bibitem[{\citenamefont{Chen}(2019)}]{chen2019exact3d}
\bibinfo{author}{\bibfnamefont{Y.-A.} \bibnamefont{Chen}},
  \bibinfo{journal}{arXiv preprint arXiv:1911.00017}  (\bibinfo{year}{2019}).

\bibitem[{\citenamefont{Shukla et~al.}(2020)\citenamefont{Shukla, Ellison, and
  Fidkowski}}]{shukla2020tensor}
\bibinfo{author}{\bibfnamefont{S.~K.} \bibnamefont{Shukla}},
  \bibinfo{author}{\bibfnamefont{T.~D.} \bibnamefont{Ellison}},
  \bibnamefont{and}
  \bibinfo{author}{\bibfnamefont{L.}~\bibnamefont{Fidkowski}},
  \bibinfo{journal}{Physical Review B} \textbf{\bibinfo{volume}{101}},
  \bibinfo{pages}{155105} (\bibinfo{year}{2020}).

\bibitem[{\citenamefont{Tantivasadakarn}(2020)}]{tantivasadakarn2020jordan}
\bibinfo{author}{\bibfnamefont{N.}~\bibnamefont{Tantivasadakarn}},
  \bibinfo{journal}{Physical Review Research} \textbf{\bibinfo{volume}{2}},
  \bibinfo{pages}{023353} (\bibinfo{year}{2020}).

\bibitem[{\citenamefont{Kitaev}(2003)}]{kitaev2003fault}
\bibinfo{author}{\bibfnamefont{A.~Y.} \bibnamefont{Kitaev}},
  \bibinfo{journal}{Annals of Physics} \textbf{\bibinfo{volume}{303}},
  \bibinfo{pages}{2} (\bibinfo{year}{2003}).

\bibitem[{\citenamefont{Levin and Wen}(2003)}]{levin2003fermions}
\bibinfo{author}{\bibfnamefont{M.}~\bibnamefont{Levin}} \bibnamefont{and}
  \bibinfo{author}{\bibfnamefont{X.-G.} \bibnamefont{Wen}},
  \bibinfo{journal}{Physical Review B} \textbf{\bibinfo{volume}{67}},
  \bibinfo{pages}{245316} (\bibinfo{year}{2003}).

\bibitem[{\citenamefont{Levin and Wen}(2005)}]{levin2005string}
\bibinfo{author}{\bibfnamefont{M.~A.} \bibnamefont{Levin}} \bibnamefont{and}
  \bibinfo{author}{\bibfnamefont{X.-G.} \bibnamefont{Wen}},
  \bibinfo{journal}{Physical Review B} \textbf{\bibinfo{volume}{71}},
  \bibinfo{pages}{045110} (\bibinfo{year}{2005}).

\bibitem[{\citenamefont{Wen}(2003)}]{wen2003quantum}
\bibinfo{author}{\bibfnamefont{X.-G.} \bibnamefont{Wen}},
  \bibinfo{journal}{Physical review letters} \textbf{\bibinfo{volume}{90}},
  \bibinfo{pages}{016803} (\bibinfo{year}{2003}).

\bibitem[{\citenamefont{Nussinov et~al.}(2012)\citenamefont{Nussinov, Ortiz,
  and Cobanera}}]{majorana_dualities}
\bibinfo{author}{\bibfnamefont{Z.}~\bibnamefont{Nussinov}},
  \bibinfo{author}{\bibfnamefont{G.}~\bibnamefont{Ortiz}}, \bibnamefont{and}
  \bibinfo{author}{\bibfnamefont{E.}~\bibnamefont{Cobanera}},
  \bibinfo{journal}{Physical Review B} \textbf{\bibinfo{volume}{86}},
  \bibinfo{pages}{085415} (\bibinfo{year}{2012}).

\bibitem[{\citenamefont{Zohar and Cirac}(2018)}]{zohar2018eliminating}
\bibinfo{author}{\bibfnamefont{E.}~\bibnamefont{Zohar}} \bibnamefont{and}
  \bibinfo{author}{\bibfnamefont{J.~I.} \bibnamefont{Cirac}},
  \bibinfo{journal}{Physical Review B} \textbf{\bibinfo{volume}{98}},
  \bibinfo{pages}{075119} (\bibinfo{year}{2018}).

\bibitem[{\citenamefont{Sachdev and Ye}(1993)}]{SY_model}
\bibinfo{author}{\bibfnamefont{S.}~\bibnamefont{Sachdev}} \bibnamefont{and}
  \bibinfo{author}{\bibfnamefont{J.}~\bibnamefont{Ye}},
  \bibinfo{journal}{Physical Review Letters} \textbf{\bibinfo{volume}{70}},
  \bibinfo{pages}{3339} (\bibinfo{year}{1993}).

\bibitem[{\citenamefont{Sachdev}(2015)}]{Sachdev_blackhole}
\bibinfo{author}{\bibfnamefont{S.}~\bibnamefont{Sachdev}},
  \bibinfo{journal}{Physical Review X} \textbf{\bibinfo{volume}{5}},
  \bibinfo{pages}{041025} (\bibinfo{year}{2015}).

\bibitem[{\citenamefont{Garc{\'\i}a-{\'A}lvarez
  et~al.}(2017)\citenamefont{Garc{\'\i}a-{\'A}lvarez, Egusquiza, Lamata,
  Del~Campo, Sonner, and Solano}}]{garcia2017digitalSYK}
\bibinfo{author}{\bibfnamefont{L.}~\bibnamefont{Garc{\'\i}a-{\'A}lvarez}},
  \bibinfo{author}{\bibfnamefont{I.}~\bibnamefont{Egusquiza}},
  \bibinfo{author}{\bibfnamefont{L.}~\bibnamefont{Lamata}},
  \bibinfo{author}{\bibfnamefont{A.}~\bibnamefont{Del~Campo}},
  \bibinfo{author}{\bibfnamefont{J.}~\bibnamefont{Sonner}}, \bibnamefont{and}
  \bibinfo{author}{\bibfnamefont{E.}~\bibnamefont{Solano}},
  \bibinfo{journal}{Physical Review Letters} \textbf{\bibinfo{volume}{119}},
  \bibinfo{pages}{040501} (\bibinfo{year}{2017}).

\bibitem[{\citenamefont{Babbush et~al.}(2019)\citenamefont{Babbush, Berry, and
  Neven}}]{babbush2019quantumSYK}
\bibinfo{author}{\bibfnamefont{R.}~\bibnamefont{Babbush}},
  \bibinfo{author}{\bibfnamefont{D.~W.} \bibnamefont{Berry}}, \bibnamefont{and}
  \bibinfo{author}{\bibfnamefont{H.}~\bibnamefont{Neven}},
  \bibinfo{journal}{Physical Review A} \textbf{\bibinfo{volume}{99}},
  \bibinfo{pages}{040301} (\bibinfo{year}{2019}).

\bibitem[{\citenamefont{Cao et~al.}(2020)\citenamefont{Cao, Zhou, Shi, and
  Zhang}}]{cao2020towardsSYK}
\bibinfo{author}{\bibfnamefont{Y.}~\bibnamefont{Cao}},
  \bibinfo{author}{\bibfnamefont{Y.-N.} \bibnamefont{Zhou}},
  \bibinfo{author}{\bibfnamefont{T.-T.} \bibnamefont{Shi}}, \bibnamefont{and}
  \bibinfo{author}{\bibfnamefont{W.}~\bibnamefont{Zhang}},
  \bibinfo{journal}{arXiv preprint arXiv:2003.01514}  (\bibinfo{year}{2020}).

\bibitem[{\citenamefont{Vidal}(2004)}]{vidal_1D}
\bibinfo{author}{\bibfnamefont{G.}~\bibnamefont{Vidal}},
  \bibinfo{journal}{Physical Review Letters} \textbf{\bibinfo{volume}{93}},
  \bibinfo{pages}{040502} (\bibinfo{year}{2004}).

\bibitem[{\citenamefont{Swingle}(2012)}]{swingle2012entanglement}
\bibinfo{author}{\bibfnamefont{B.}~\bibnamefont{Swingle}},
  \bibinfo{journal}{Physical Review D} \textbf{\bibinfo{volume}{86}},
  \bibinfo{pages}{065007} (\bibinfo{year}{2012}).

\bibitem[{\citenamefont{Pastawski et~al.}(2015)\citenamefont{Pastawski,
  Yoshida, Harlow, and Preskill}}]{pastawski2015holographic}
\bibinfo{author}{\bibfnamefont{F.}~\bibnamefont{Pastawski}},
  \bibinfo{author}{\bibfnamefont{B.}~\bibnamefont{Yoshida}},
  \bibinfo{author}{\bibfnamefont{D.}~\bibnamefont{Harlow}}, \bibnamefont{and}
  \bibinfo{author}{\bibfnamefont{J.}~\bibnamefont{Preskill}},
  \bibinfo{journal}{Journal of High Energy Physics}
  \textbf{\bibinfo{volume}{2015}}, \bibinfo{pages}{149} (\bibinfo{year}{2015}).

\bibitem[{\citenamefont{Xu et~al.}(2020)\citenamefont{Xu, Susskind, Su, and
  Swingle}}]{xu2020sparseSYK}
\bibinfo{author}{\bibfnamefont{S.}~\bibnamefont{Xu}},
  \bibinfo{author}{\bibfnamefont{L.}~\bibnamefont{Susskind}},
  \bibinfo{author}{\bibfnamefont{Y.}~\bibnamefont{Su}}, \bibnamefont{and}
  \bibinfo{author}{\bibfnamefont{B.}~\bibnamefont{Swingle}},
  \bibinfo{journal}{arXiv preprint arXiv:2008.02303}  (\bibinfo{year}{2020}).

\bibitem[{\citenamefont{Chen et~al.}(2010)\citenamefont{Chen, Gu, and
  Wen}}]{chen2010local}
\bibinfo{author}{\bibfnamefont{X.}~\bibnamefont{Chen}},
  \bibinfo{author}{\bibfnamefont{Z.-C.} \bibnamefont{Gu}}, \bibnamefont{and}
  \bibinfo{author}{\bibfnamefont{X.-G.} \bibnamefont{Wen}},
  \bibinfo{journal}{Physical Review B} \textbf{\bibinfo{volume}{82}},
  \bibinfo{pages}{155138} (\bibinfo{year}{2010}).

\bibitem[{\citenamefont{Aguado and Vidal}(2008)}]{aguado2008entanglement}
\bibinfo{author}{\bibfnamefont{M.}~\bibnamefont{Aguado}} \bibnamefont{and}
  \bibinfo{author}{\bibfnamefont{G.}~\bibnamefont{Vidal}},
  \bibinfo{journal}{Physical Review Letters} \textbf{\bibinfo{volume}{100}},
  \bibinfo{pages}{070404} (\bibinfo{year}{2008}).

\bibitem[{\citenamefont{Bravyi et~al.}(2006)\citenamefont{Bravyi, Hastings, and
  Verstraete}}]{bravyi2006lieb}
\bibinfo{author}{\bibfnamefont{S.}~\bibnamefont{Bravyi}},
  \bibinfo{author}{\bibfnamefont{M.~B.} \bibnamefont{Hastings}},
  \bibnamefont{and}
  \bibinfo{author}{\bibfnamefont{F.}~\bibnamefont{Verstraete}},
  \bibinfo{journal}{Physical Review Letters} \textbf{\bibinfo{volume}{97}},
  \bibinfo{pages}{050401} (\bibinfo{year}{2006}).

\bibitem[{\citenamefont{Gu et~al.}(2017)\citenamefont{Gu, Qi, and
  Stanford}}]{gu2017latticeSYK}
\bibinfo{author}{\bibfnamefont{Y.}~\bibnamefont{Gu}},
  \bibinfo{author}{\bibfnamefont{X.-L.} \bibnamefont{Qi}}, \bibnamefont{and}
  \bibinfo{author}{\bibfnamefont{D.}~\bibnamefont{Stanford}},
  \bibinfo{journal}{Journal of High Energy Physics}
  \textbf{\bibinfo{volume}{2017}}, \bibinfo{pages}{125} (\bibinfo{year}{2017}).

\bibitem[{\citenamefont{Song et~al.}(2017)\citenamefont{Song, Jian, and
  Balents}}]{song2017latticeSYK}
\bibinfo{author}{\bibfnamefont{X.-Y.} \bibnamefont{Song}},
  \bibinfo{author}{\bibfnamefont{C.-M.} \bibnamefont{Jian}}, \bibnamefont{and}
  \bibinfo{author}{\bibfnamefont{L.}~\bibnamefont{Balents}},
  \bibinfo{journal}{Physical Review Letters} \textbf{\bibinfo{volume}{119}},
  \bibinfo{pages}{216601} (\bibinfo{year}{2017}).

\bibitem[{\citenamefont{Chamberland et~al.}(2020)\citenamefont{Chamberland,
  Zhu, Yoder, Hertzberg, and Cross}}]{chamberland2020topological}
\bibinfo{author}{\bibfnamefont{C.}~\bibnamefont{Chamberland}},
  \bibinfo{author}{\bibfnamefont{G.}~\bibnamefont{Zhu}},
  \bibinfo{author}{\bibfnamefont{T.~J.} \bibnamefont{Yoder}},
  \bibinfo{author}{\bibfnamefont{J.~B.} \bibnamefont{Hertzberg}},
  \bibnamefont{and} \bibinfo{author}{\bibfnamefont{A.~W.} \bibnamefont{Cross}},
  \bibinfo{journal}{Physical Review X} \textbf{\bibinfo{volume}{10}},
  \bibinfo{pages}{011022} (\bibinfo{year}{2020}).

\bibitem[{\citenamefont{Arute et~al.}(2019)\citenamefont{Arute, Arya, Babbush,
  Bacon, Bardin, Barends, Biswas, Boixo, Brandao, Buell
  et~al.}}]{arute2019quantumsupremacy}
\bibinfo{author}{\bibfnamefont{F.}~\bibnamefont{Arute}},
  \bibinfo{author}{\bibfnamefont{K.}~\bibnamefont{Arya}},
  \bibinfo{author}{\bibfnamefont{R.}~\bibnamefont{Babbush}},
  \bibinfo{author}{\bibfnamefont{D.}~\bibnamefont{Bacon}},
  \bibinfo{author}{\bibfnamefont{J.~C.} \bibnamefont{Bardin}},
  \bibinfo{author}{\bibfnamefont{R.}~\bibnamefont{Barends}},
  \bibinfo{author}{\bibfnamefont{R.}~\bibnamefont{Biswas}},
  \bibinfo{author}{\bibfnamefont{S.}~\bibnamefont{Boixo}},
  \bibinfo{author}{\bibfnamefont{F.~G.} \bibnamefont{Brandao}},
  \bibinfo{author}{\bibfnamefont{D.~A.} \bibnamefont{Buell}},
  \bibnamefont{et~al.}, \bibinfo{journal}{Nature}
  \textbf{\bibinfo{volume}{574}}, \bibinfo{pages}{505} (\bibinfo{year}{2019}).

\bibitem[{\citenamefont{Cross et~al.}(2019)\citenamefont{Cross, Bishop,
  Sheldon, Nation, and Gambetta}}]{cross2019validating}
\bibinfo{author}{\bibfnamefont{A.~W.} \bibnamefont{Cross}},
  \bibinfo{author}{\bibfnamefont{L.~S.} \bibnamefont{Bishop}},
  \bibinfo{author}{\bibfnamefont{S.}~\bibnamefont{Sheldon}},
  \bibinfo{author}{\bibfnamefont{P.~D.} \bibnamefont{Nation}},
  \bibnamefont{and} \bibinfo{author}{\bibfnamefont{J.~M.}
  \bibnamefont{Gambetta}}, \bibinfo{journal}{Physical Review A}
  \textbf{\bibinfo{volume}{100}}, \bibinfo{pages}{032328}
  (\bibinfo{year}{2019}).

\end{thebibliography}

\end{document}